%% file: main_SM_for_arxiv.tex
\documentclass[twocolumn,
amsmath,amssymb,
 aps,
 prl,
 floatfix,
 superscriptaddress, 
 nofootinbib
]{revtex4}
\usepackage{tabularray}
\usepackage{graphicx}
\usepackage{dcolumn}
\usepackage{bm}
\usepackage{dsfont}

\usepackage[normalem]{ulem} 

\usepackage[center]{caption}
\usepackage{xcolor}
\usepackage{hyperref}
\usepackage{cleveref}
\usepackage{gensymb}

\begin{document}

\input{main_for_arxiv.tex}

\clearpage

\widetext
\input{SM_for_arxiv.tex}

\bibliographystyle{apsrev4-1}
\bibliography{main_SM_for_arxiv}

\end{document}

%% file: main_for_arxiv.tex

 \title{
Search for QCD coupled axion dark matter with the MICROSCOPE space experiment}

\author{Jordan Gu\'e}\email{jgue@ifae.es}
\affiliation{Institut de F\'isica d’Altes Energies (IFAE), The Barcelona Institute of Science and Technology, Campus UAB, 08193 Bellaterra (Barcelona), Spain}
\author{Peter Wolf}
\affiliation{%
 LTE, Observatoire de Paris, Université PSL, Sorbonne Université, Université de Lille, LNE, CNRS, 61 Avenue de l’Observatoire, 75014 Paris, France
}
\author{Aur\'elien Hees}
\affiliation{%
 LTE, Observatoire de Paris, Université PSL, Sorbonne Université, Université de Lille, LNE, CNRS, 61 Avenue de l’Observatoire, 75014 Paris, France
}
%

\begin{abstract}
Axion dark matter coupled via QCD induces a non-zero differential acceleration between test masses of different composition. Tests of the equivalence principle, like the recent MICROSCOPE space mission, are sensitive to such a signal. We use the final released data of the MICROSCOPE experiment, to search for this effect. We find no positive signal consistent with the dark matter model, and set upper limits on the axion-gluon coupling that improve existing laboratory bounds by up to two orders of magnitude for axion masses in the $10^{-17}$~eV to $10^{-13}$~eV range.
\end{abstract}

\maketitle

\section{\label{sec:Intro}Introduction}

Almost 100 years after its first observational evidence \cite{Zwicky33}, the fundamental nature of dark matter (DM) is still to this day one of the biggest mysteries in fundamental physics \cite{Bertone18}. Out of all the possible DM candidates, ultralight fields (whose mass is below the eV) are a compelling alternative to historically dominant models, such as WIMPs (Weakly Interacting Massive Particles). In these models, DM is described by a non relativistic classical wave oscillating at its Compton frequency \cite{Hui21,Cirelli24}. One of the most prominent ultralight dark matter candidate (ULDM) is the QCD axion, originally postulated fifty years ago to solve the so-called strong CP problem \cite{Peccei77}. More general pseudo-scalars, more commonly referred to as axion-like particles (ALP), which do not solve the strong CP problem, can still account for the galactic DM, and therefore are actively searched for \cite{marsh:2016aa, OHare24,AxionLimits}. 

In the simplest scenario, the mixing between the (QCD) axion and neutral pions (with strength $1/f_a$) naturally solve the strong CP problem, leading to the so-called linear relationship between the axion mass and $1/f_a$ \cite{diCortona16,Weinberg78}. However, some models allow for such mixing without the need of this relation, i.e at fixed coupling value, lower mass axions are feasible \cite{DiLuzio21,Hook18-2}.

As a consequence of the QCD-axion coupling, the axion potential becomes dependent on the light quarks and pion masses after the QCD phase transition \cite{di-vecchia:1980aa,diCortona16}. Equivalently, this also leads to a quadratic dependence of the pion masses on the axion field, see e.g. \cite{diCortona16,Kim22}. Among other observables, this can lead to very strong forces mediated by the axion between very dense astrophysical objects, such as neutron stars \cite{Hook18}, or a change of the composition of white dwarfs \cite{balkin24}.

The mass of the nucleons depends on the pion mass and this dependence has been computed at one-loop level, see e.g. \cite{gasser:1988aa,bernard:1992aa,scherer:2012aa,Kim22}. Similarly, the nucleon $g$-factor at the chiral order also depends on the pion mass \cite{scherer:2012aa,Kim22} such that atomic transition frequencies are dependent on the axion field \cite{Kim22}. Finally, the atomic nuclear binding energy at chiral order is also dependent on the pion mass, see \cite{damour:2010zr} for a detailed derivation. As a consequence, the atomic rest mass becomes quadratically dependent on the axion field~\cite{Gue:2024onx,Bauer24}. This relationship is non-universal in the sense that it depends on the atomic number and on the atomic mass, which leads to a breaking of the universality of free fall (UFF) and a violation of the equivalence principle \cite{Gue:2024onx}.

At cosmological scales, the axion oscillates in its self-potential at its Compton frequency \cite{marsh:2016aa} and these oscillations may comprise DM. In this scenario, the oscillations amplitude is stochastically related to the DM energy density and the pion masses, atomic frequencies and atomic rest-masses all become time dependent \cite{diCortona16,Kim22,Gue:2024onx}.  These oscillations have been searched for in several atomic oscillators \cite{Madge24, Zhang23}, but recently, it was suggested that they can also be searched for in classical UFF tests  or using atom interferometers \cite{Gue:2024onx}. 

In addition, the quadratic axion-pion coupling induces more complex signatures similarly to the ones arising for quadratically coupled pure scalar fields \cite{hees18}. In particular, in the vicinity of massive bodies, the back action of matter on the axion field can strongly impact its oscillation amplitude, a property which was originally derived under the spherical symmetry assumption in \cite{hees18} where the small field gradient was neglected. Recently, those solutions have been extended to account for the field propagation \cite{Banerjee25,DelCastillo25}.

The matter-axion coupling being quadratic implies that in the presence of a massive body the breaking of the UFF induced by an ALP has two distinct signatures \cite{hees18}:  (i) oscillations at twice the axion Compton frequency and (ii) a static differential acceleration directed towards the centre of the massive body, which corresponds to the one searched for in standard UFF tests. Currently, the best UFF test is provided by the MICROSCOPE experiment  \cite{Microscope22,Touboul22cqg}, a space-based experiment which monitored the differential acceleration between two test masses made of Platinum and Titanium and which provided a constraint on the E\"otvos parameter at the level of $10^{-15}$ \cite{Microscope22}. 

In this Letter, we reanalyze the publicly available MICROSCOPE data \cite{Microscope22,Touboul22cqg} to search for axion DM through its coupling to gluons. To do so, we re-combine the individual measurement session results given in \cite{Touboul22cqg} in a way that is optimized for our DM search. We find no evidence of axion DM coupled to gluons, and this allows us to set a new upper bound on the coupling strength $1/f_a$. Our constraint covers seven orders of magnitude in mass from $10^{-20}$ to $\sim 10^{-13}$ eV, and improves current laboratory bounds by up to two orders of magnitude for axion masses between $10^{-17}$ and $\sim 10^{-13}$ eV, reaching $1/f_a \sim 10^{-16}$ GeV$^{-1}$ at $10^{-20}$ eV. 

\section{Breaking of the universality of free fall by an axion field}\label{sec:theory}
As mentioned above, the axion-gluon coupling leads to a quadratic dependence of the pion mass on the axion field and subsequently to a violation of the universality of free fall. We will present the theoretical modeling of the axion induced UFF breaking. A detailed derivation and discussion is presented in Sec.~A of the Supplemental Material. 

We start from the Lagrangian\footnote{We use a convention where the axion field amplitude is dimensionless. Other conventions exist, see e.g. \cite{Kim22,Gue:2024onx}.}
\begin{align}
   \mathcal{L} &= -\frac{1}{2\kappa}g^{\mu\nu}\partial_\mu a \partial_\nu a -\frac{m^2_a c^2}{2\hbar^2 \kappa}a^2 + E_P\frac{g^2_3}{32\pi^2}\frac{a}{f_a} G^a_{\mu\nu}\tilde G^{a,\mu\nu} \, ,
\end{align}
where $\kappa=8\pi G/c^4$ is the Einstein gravitational constant, $g_{\mu\nu}$ is the spacetime metric, $a$ is the dimensionless ALP of mass $m_a$, $E_P=\sqrt{\hbar c/\kappa}$ is the reduced Planck energy, $f_a$ is the ALP-gluon coupling (which corresponds to the Peccei-Quinn spontaneous symmetry breaking scale \cite{Peccei77}), $G_{\mu\nu}$ is the QCD strength tensor and $g_3$ is the strong coupling.

This Lagrangian leads to a mixing between the ALP and the neutral pion, as both are pseudo-Goldstone bosons, such that the pion mass $m_\pi$ becomes quadratically dependent on $\theta\equiv E_P\: a/f_a$ \cite{diCortona16,Kim22}
\begin{equation}
    \frac{\delta m_\pi^2}{m_\pi^2} =- \frac{m_u m_d}{2\left(m_u+m_d\right)^2}\theta^2 \, ,
\end{equation}
where $m_{u/d}$ are the up/down quarks masses. In the chiral perturbation theory, the mass of the nucleons \cite{gasser:1988aa,bernard:1992aa,scherer:2012aa,Kim22} and the nuclear binding energy \cite{damour:2010zr} are both dependent on the pion mass. When combined together, this implies that the rest mass of an atom $A$ depends quadratically on the ALP field, with an atom-dependent coefficient, the axionic charge $Q^A_M$, i.e
\begin{align}
    m_A &= m^0_A\left(1+Q^A_M\frac{ E^2_P}{f^2_a}a^2\right) \, ,
\end{align}
where $m^0_A$ is the unperturbed mass.  These charges depend on the atomic mass number $A$ and charge number $Z$ and they are derived explicitly in \cite{Gue:2024onx}\footnote{\label{fn:charges}Relevant to this analysis, the axionic charge of the Earth, modelled as a homogeneous SiO$_2$ ball, is $Q_M^\mathrm{E}=-68.442 \times 10^{-3}$ and for the two MICROSCOPE test masses $Q_M^\mathrm{Pt}=-69.065\times 10^{-3}$ and $Q_M^\mathrm{Ti}=-68.770\times 10^{-3}$.}. The non universality of $Q^A_M$ induces a violation of the UFF \cite{Gue:2024onx}, which in the case of quadratic couplings, leads to a differential acceleration between two co-located test masses $A$ and $B$ \cite{hees18} 
\begin{equation}\label{eq:delta_a}
    \Delta \vec a = -2 \left(Q_M^A-Q_M^B\right) \left(\frac{E_P}{f_a}\right)^2 a\left(c^2 \vec\nabla a + \vec v \dot a\right)\, .
\end{equation}

At cosmological scales, the axion oscillates around the minimum of its potential at its Compton frequency $\omega_a = m_ac^2/\hbar$ \cite{marsh:2016aa}. If the axion constitutes DM, its oscillation amplitude $a_0$ depends on the DM energy density through $a_0 = \sqrt{16\pi G \rho_\mathrm{DM}}/\omega_a c$ with $\rho_\mathrm{DM}=0.4$ GeV/cm$^3$ \cite{McMillan11}. Locally, close to massive bodies, the back action of matter on the axion field can strongly impact its oscillation amplitude which becomes location dependent, i.e.
\begin{subequations}\label{eq:sol}
\begin{equation}\label{eq:axion_sol}
    a(t,\vec r) =  a_0 A(\vec r)\cos(\omega_a t + \Phi(r)) \, ,
\end{equation}
where $\Phi(r)$ is a phase, which can be position dependent as well (see below and SM). 

The solution $A(\vec r)$ shows a scalarization mechanism (i.e. $A(r)$ can become large close to massive bodies) \cite{hees18}. Neglecting (i) the backreaction from the metric and (ii) the small contribution from the axion asymptotic gradient $k\sim \omega_a v_\mathrm{DM}/c^2\sim 10^{-3}\omega_a/c$ (with $v_\mathrm{DM}$ the DM velocity in the Solar System frame), this can  lead to divergences of the amplitude for certain values of the coupling parameter \cite{hees18}. 

Recently, these solutions have been extended to account for $k$ under the form of a multipole extension \cite{Banerjee25,DelCastillo25}. As pointed out in \cite{Banerjee25,DelCastillo25}, one important contribution of the gradient is to regularize the divergences of the original solution which become resonances.  Note that the monopole contribution is the dominant one by a factor $1/kR_E \gtrapprox 10^3\gg 1$ (see SM) such that we consider only this one in our data analysis, i.e. the $A(r)$ function writes
\begin{align}
    A(r) &= \frac{\left|\lambda\cos\lambda\sin(kr-\beta)+\beta\cos(kr-\beta)\sin\lambda\right|}{kr\sqrt{\lambda^2\cos^2\lambda+\beta^2\sin^2\lambda}}\, ,
\end{align}
where 
\begin{align}
    \beta = kR_E \: , \: \lambda = \sqrt{\beta^2+12\left(\frac{E_P}{f_a}\right)^2|Q^E_M|\frac{G M_E}{R_Ec^2}}\, ,
\end{align}
\end{subequations}
where $R_E$ is the Earth radius modeled as a spherically homogeneous body of mass $M_E$ and of charge $Q^E_M$, see \cref{fn:charges}. 

\begin{figure}
    \centering
    \includegraphics[width=0.4\textwidth]{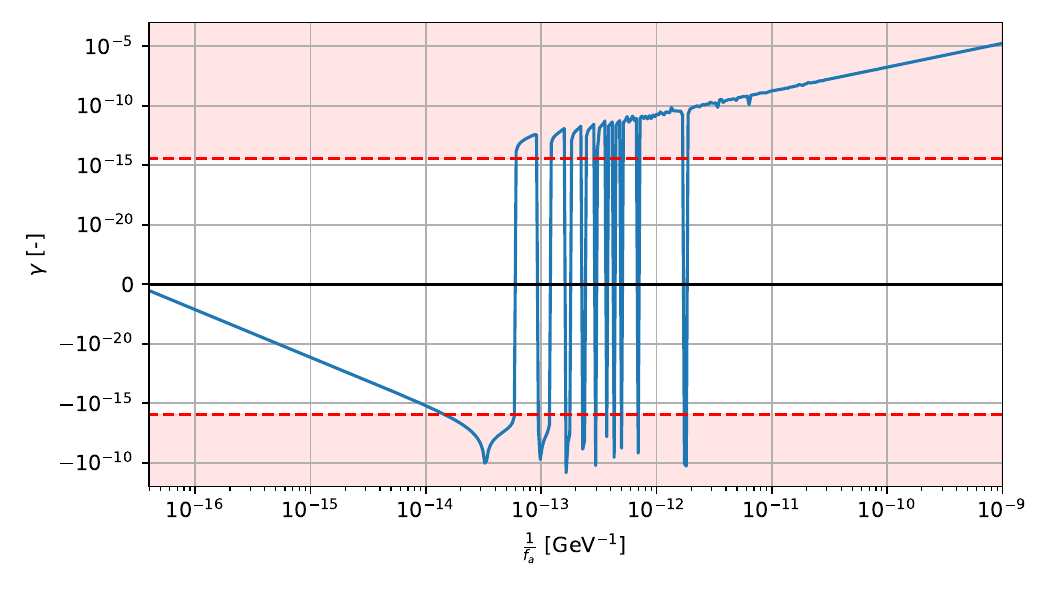}
    \caption{In blue the evolution of the signal strength $\gamma$ (c.f. Eq.~\eqref{eq:eta}) as a function of the coupling $1/f_a$ for one value of the axion mass ($m_a=10^{-16}$~eV). The red dashed lines correspond to the lower and upper limits from the MICROSCOPE data analysis, the red filled surface corresponding to the exclusion area.}
    \label{fig:axion_gamma}
\end{figure}

In standard models of galaxy formation, galactic DM must be virialized \cite{freese:2013aa,pillepich:2014aa} such that its velocity distribution acquires a characteristic width $\sigma_v\sim 10^{-3}c$ \cite{Cirelli24,Evans19}, and the DM oscillations have a characteristic coherence time $\tau \sim c^2/\sigma^2_v \times 2\pi \hbar/(m_a c^2)$\cite{Derevianko18}.
This implies that for observation times longer than $\tau$, the DM spectral signature is broadened \cite{foster:2018aa,Centers21}. For observation times shorter than $\tau$, the signal is effectively monochromatic, but its amplitude receives a stochastic correction \cite{foster:2018aa,Centers21}. In practice, one needs to introduce a stochastic parameter $\alpha$, which follows a Rayleigh distribution such that the amplitude of the axion wave becomes proportional to $\alpha$ \cite{foster:2018aa,Centers21,Savalle21,nakatsuka:2023aa}.

The axion field of Eq.~(\ref{eq:axion_sol}) induces two types of signatures on the UFF signal of Eq.~(\ref{eq:delta_a}): (i) a fast oscillating signal (at frequency $2\omega_a$) and (ii) a static position dependent signal. In this Letter, we focus on the static signal which gives a relative differential acceleration between two test masses of
\begin{equation}\label{eq:eta}
    \eta = \frac{\left|\Delta \vec a\right|}{\left| \vec a\right|} = \alpha^2 \Delta Q_M \frac{c^2a_0^2}{2\frac{GM_E}{r^2}}\left(\frac{E_P}{f_a}\right)^2 \frac{\partial A^2(r)}{\partial r}\equiv \gamma \alpha^2 \, ,
\end{equation}
where $\eta$ is the E\"otvos parameter, $\Delta Q_M = (Q^B_M-Q^A_M)$, and $\alpha^2$, which follows an exponential distribution, is due to the DM velocity distribution. The $\gamma$ function, which characterizes the strength of the signal searched for in the data, is depicted for one axion mass on Fig.~\ref{fig:axion_gamma} (see the Supplemental material for more details).


\section{MICROSCOPE data}\label{sec:micro}

MICROSCOPE is a space mission designed to probe the UFF in space at the level of $10^{-15}$ \cite{Microscope17} by measuring the differential acceleration between a test mass made of Titanium and another one made of Platinum. The two free falling bodies are concentric cylindrical test masses
controlled in a differential accelerometer. In total, 18 measurement sessions taken between 2016 and 2018  have been used to constrain the E\"otvos parameter \cite{Touboul22cqg}. The duration of a session lasts between 13 hours and 8 days. Each session has been analyzed independently providing an estimate of the E\"otvos parameter $\eta_i$, a statistical error $\sigma_{\mathrm{stat}, i}$ and a systematic error $\sigma_{\mathrm{syst}, i}$ \cite{Touboul22cqg}, which are provided in the Supplemental Material. Note that in our analysis, we quadratically combine the statistical and systematic uncertainties.

\section{Bayesian parameter inference}\label{sec:bayes_analysis}
The methodology used in this analysis is thoroughly detailed in the Sec.~B of the Supplemental Material. We present here the important steps. We use a Bayesian approach and assume the measurement noise to follow a Gaussian distribution. 

For a given axion mass $m_a$, we compute the axion field coherence time and we group the MICROSCOPE sessions into subsets where the total duration spanned by a subset of sessions is smaller than the coherence time. Within each subset of sessions, we can therefore use the monochromatic modeling of the axion field discussed above. Note that the $\alpha$ stochastic parameter is different and independent for all the subsets. Within each subset of sessions, we combine the individual likelihoods and subsequently marginalize over their $\alpha$ parameter. The combined likelihood of all the measurements is simply provided by the product of the marginalized likelihood from all the subsets.

Following \cite{Centers21}, we use a non informative Berger–Bernardo reference prior, which in our case is equivalent to Jeffrey’s prior. This non informative prior is motivated because (i) it ensures invariance of the inference results under a change of  variables, (ii)  it  ensures  that  the integral of the posterior converges  (which is not the case if one chooses naively a uniform prior on $\gamma$), (iii) it maximizes the Kullback–Leibler of the posterior with respect to the prior and (iv) it produces results consistent with the frequentist approach \cite{Centers21}. This prior is estimated numerically (see Supplemental material for more details).

The product of the marginalized likelihood with the prior allows us to estimate the posterior distribution of the $\gamma$ parameter which characterizes our signal, see Eq.~(\ref{eq:eta}). The ratio of the Bayesian evidence between this model which contain an axion signal and a model which includes only noise is then computed and used to assess if an axion signal is significantly detected within the MICROSCOPE data. 

In the case of no positive detection, we use the posterior probability distribution to infer a 95\% credible interval on $\gamma$. This interval depends on the mass of the axion field $m_a$ and is illustrated by the red dashed curves on the bottom panel from Fig.~\ref{fig:axion_gamma} in the case of $m_a=10^{-16}$ eV. The non-linear expression of $\gamma$ as a function of $1/f_a$ allows us to translate the constraint on $\gamma$ into a constraint on $1/f_a$.

This methodology is applied for different axion masses.

\section{Results and discussion}\label{sec:results}

We first assess if there is a positive DM signal in the data by computing the Bayes factor $\mathcal{B}$ as the ratio of evidences of the ALP-gluon coupled model and a model assuming no signal. Over the full mass range of interest,  $\mathcal{B}$ is $< 1$ (see SM), which indicates no significant detection of DM in the data. 

\begin{figure}
    \centering
    \includegraphics[width=0.5\textwidth]{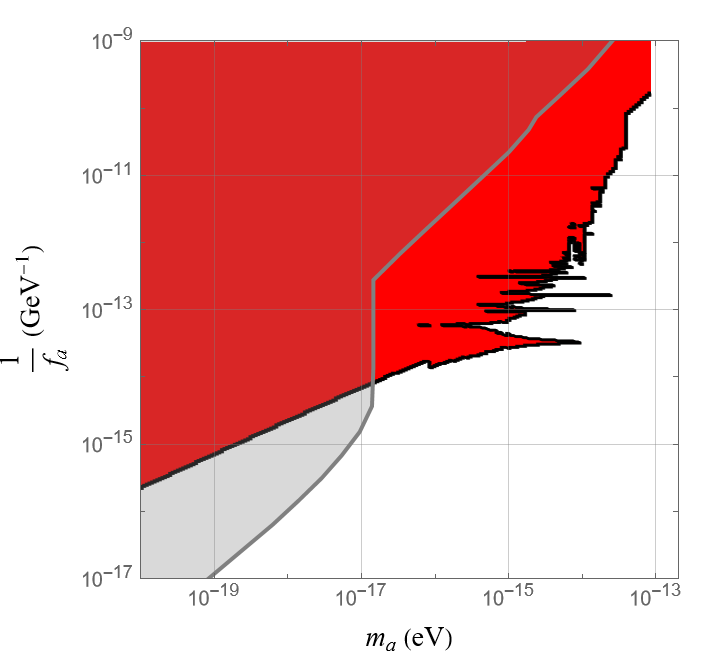}
    \caption{95\% confidence area in the ALP-gluon coupling $1/f_a$ using final results from MICROSCOPE (in red). In grey are shown the existing constraints from laboratory experiments from \cite{Beam_EDM,nEDM,Madge24} (adapted from \cite{AxionLimits}).}
    \label{fig:constraint}
\end{figure}

Since no positive signal is detected in the data, we compute an exclusion area on the value of the coupling $1/f_a$.  In Fig.~\ref{fig:constraint}, we present in red the constraint on $1/f_a$ from our analysis, while existing laboratory constraints from \cite{nEDM,Beam_EDM,Madge24} are shown by the grey-filled area. 
We find that MICROSCOPE's data constrains new regions of the parameter space, with its most stringent limit being $1/f_a \sim 1.7 \times 10^{-16}$  GeV$^{-1}$ at $m_a = 10^{-20}$ eV and reaching $1/f_a \sim 1.3 \times 10^{-10}$  GeV$^{-1}$ at $m_a \sim 7.7 \times 10^{-14}$ eV, improving by roughly two orders of magnitude the existing laboratory bounds over four orders of magnitude of mass.

The constraint of MICROSCOPE depicted in Fig.~\ref{fig:constraint} presents several bands. As shown in Fig.~\ref{fig:axion_gamma}, the detector is blind to very narrow intervals of values of couplings, around when $\gamma$ crosses zero. As discussed in the supplemental material, this happens when the axion field reaches an extremum, i.e. when its derivative changes sign.\footnote{Not all such bands are visible on Fig.~\ref{fig:constraint} because its resolution is too low to see the narrower ones (cf. Fig.~\ref{fig:axion_gamma}).}
In addition, as the axion mass increases, its coherence time decreases, such that less and less MICROSCOPE subsets of data can be constructed with the requirements $T^\mathrm{subset}_\mathrm{obs} < \tau$. As a consequence, the constraint derived from our data analysis decreases. 

In conclusion, the ALP-gluon coupling $1/f_a$ leads to a static violation of the UFF, and classical tests like MICROSCOPE can be used to constrain such coupling. In this Letter, we performed a reanalysis of MICROSCOPE final data, and we have seen no evidence for existence of an axion-gluon coupling for ALP masses between $10^{-20}$ and $10^{-13}$ eV which allows us to derive an exclusion area on its value in this corresponding mass range. Our results improve existing laboratory bounds by up to two orders of magnitude, over four orders of magnitude of mass, and can be complementary to astrophysical and cosmological bounds \cite{Hook18,Zhang21,Blum14}.

We thank the CNES and ONERA teams for making available the MICROSCOPE result, and in particular Joël Bergé, Gilles Métris and Manuel Rodrigues. This work was supported by the Programme National GRAM of CNRS/INSU with INP and IN2P3 cofunded by CNES. IFAE is partially funded by the CERCA program of the Generalitat de Catalunya. J.G. is funded by the grant CNS2023-143767. Grant CNS2023-143767 is funded by MICIU/AEI/10.13039/501100011033 and by
European Union NextGenerationEU/PRTR.

%% file: SM_for_arxiv.tex
\begin{center}
    {\large \textbf{Supplemental material: Signal modeling and data analysis}} \\[10pt]  
\end{center}

In this supplemental material, we present a detailed derivation and discussion of the signal modeling and a detailed presentation of our data analysis methodology.

\section{Signal modeling}\label{app:model}

\subsection{Axion action and field equation}

We start from the Lagrangian
\begin{align}\label{eq:lagrangian}
   \mathcal{L} &= -\frac{1}{2\kappa}g^{\mu\nu}\partial_\mu a \partial_\nu a -\frac{m^2_a c^2}{2\hbar^2 \kappa}a^2 + E_P\frac{g^2_3}{32\pi^2}\frac{a}{f_a} G^a_{\mu\nu}\tilde G^{a,\mu\nu} \, ,
\end{align}
where $a$ is the dimensionless axion of mass $m_a$, $\hbar$ is the reduced Planck constant, $E_P$ is the reduced Planck energy, $f_a$ is the axion-gluon coupling (Peccei-Quinn spontaneous symmetry breaking scale), $G_{\mu\nu}$ is the gluon strength tensor, $g_3$ is the strong coupling, and $\kappa=8\pi G/c^4$ is the Einstein gravitational constant.
Following \cite{hees18}, we derive the field equations for the axion following this Lagrangian and we get \cite{hees18}
\begin{subequations}
\begin{align}\label{eq:field_eq}
    \square_g a = - \kappa \sigma + \frac{m^2_ac^2}{\hbar^2}a \, ,
\end{align}
with $\square_g a=g^{\mu\nu}\nabla_\mu\nabla_\nu a = \frac{1}{\sqrt{-g}}\partial_\mu (\sqrt{-g}\partial^\mu a)$ and
\begin{equation}\label{eq:sigma_def}
    \sigma = \frac{1}{\sqrt{-g}} \frac{\delta \sqrt{-g} \mathcal L_\mathrm{mat}}{\delta a}=\frac{\partial \mathcal{L}_\mathrm{int}}{\partial a} \, ,
\end{equation}
 where $\mathcal{L}_\mathrm{mat}$ is the Lagrangian from matter and $\mathcal{L}_\mathrm{int}$ refers to the axion-gluon coupling term of Eq.~\eqref{eq:lagrangian}.
 
 The metric field equation in presence of the axion field reads
\begin{align}\label{eq:Einstein_eq}
    R_{\mu\nu} &=\kappa\left(T_{\mu\nu}-\frac{1}{2}g_{\mu\nu}T^\alpha_{\:\: \alpha}\right)+\partial_\mu a \partial_\nu a + g_{\mu\nu} \frac{m^2_a c^2}{\hbar^2}a^2 \, ,
\end{align}
\end{subequations}
where $R_{\mu\nu}, T_{\mu\nu}$ are respectively the Ricci tensor and the matter stress energy tensor.
 
\subsection{Matter modeling}
Based on the seminal works by \cite{Kim22} and \cite{damour:2010zr}, in \cite{Gue:2024onx} it was shown that the action used
to model matter at the microscopic level including the axion field interaction from Eq.~\eqref{eq:lagrangian} can phenomenologically be replaced at the macroscopic level by a standard point mass action where the rest mass becomes dependent on the local axion field through 
\begin{subequations}
\begin{align}\label{eq:rest_mass_axion}
    m_A &= m^0_A\left(1+\frac{a^2 E^2_P}{f^2_a}Q^A_M\right) \, ,
\end{align}
where $Q^A_M$ is the axionic charge whose expression is given by
\begin{align}
    Q^A_M &= \frac{\partial \ln m_A}{\partial \theta^2}, \quad \mathrm{with}\quad \theta = a E_P/f_a \, .
\end{align}
\end{subequations}
These axionic charge depends on the composition of body $A$ such that it leads to a breaking of the equivalence principle and therefore of the universality of free fall, see \cite{Gue:2024onx} for a detailed discussion. Assuming an atomic species of mass number $A$ and charge number $Z$, its explicit expression is \cite{Gue:2024onx}
\begin{align}
    Q_M &\approx - 0.070 + 10^{-3} \times \left(\frac{3.98}{A^{1/3}}+ 2.22\frac{(A-2Z)^2}{A^2}+ 0.015\frac{Z(Z-1)}{A^{4/3}}\right)\, .
\end{align}
In Table ~\ref{tab:axionic_charge}, we provide the numerical values of the axionic charges of some species of interest.
\begin{table}
\resizebox{0.3\textwidth}{!}{
\begin{tabular}{|c|c|}
\hline
Atomic species & $Q_M$ ($\times 10^{-3}$)\\
\hline
 $^{195}$Pt &  -69.065\\
\hline
$^{48}$Ti & -68.770\\
\hline
SiO$_2$  & -68.442\\
 \hline
\end{tabular}}
\caption{Some axionic charges used in this analysis.}
\label{tab:axionic_charge}
\end{table}

We model standard matter as a pressureless perfect fluid whose stress-energy tensor is given by $T^{\mu\nu}= \rho u^\mu u^\nu$, where $\rho$ is the energy density and $u^\mu$ the
4-velocity of the fluid. For this matter modeling, the source
term for a body $A$ in the Klein-Gordon equation is written as \cite{hees18,Hees15}
\begin{align}\label{eq:sigma_A}
    \sigma_A &= \frac{\partial \mathcal{L}_\mathrm{int,A}}{\partial a} = -\frac{\partial \ln m_A}{\partial a}\rho_A  \, ,
\end{align}
To first order, Eq.~\eqref{eq:rest_mass_axion} leads to 
\begin{subequations}
\begin{align}\label{eq:partial_ln_ma}
   \frac{\partial \ln m_A}{\partial a} &= \frac{2 a E^2_P Q^A_M}{f^2_a} \, ,
\end{align}
such that Eq.~\eqref{eq:sigma_A} becomes
\begin{align}\label{eq:sigma}
    \sigma_A &= -\frac{2E^2_P Q^A_M \rho_A}{f^2_a} a\equiv \chi_A a \, .
\end{align}
\end{subequations}
Note that as derived in \cite{Gue:2024onx}, the axionic charges are negative, i.e $\sigma >0$. Note also that $\sigma$ vanishes in a vacuum.

\subsection{Axion field}
In this section, we will review the solution to the Klein-Gordon  Eq.~\eqref{eq:field_eq} under the assumption that the background metric is the Minkowski metric, i.e. 
\begin{equation}\label{eq:KG_lin}
    -\frac{1}{c^2}\frac{\partial^2 a}{\partial t^2}+\Delta a -\frac{m^2_ac^2}{\hbar^2}a = -\kappa \sigma\, .
\end{equation}
We will justify a posteriori this approximation.

\subsubsection{Solution in a vacuum}
In a vacuum, Eq.~\eqref{eq:field_eq} is a standard Klein-Gordon equation whose solutions are standard plane wave 
\begin{equation}\label{eq:free_sol}
a_\mathrm{free}(t,\vec x) = a_0 \cos(\omega_a t-\vec k \cdot \vec x+\Phi) \, ,
\end{equation}
where $\Phi$ a constant phase, $\vec k$ the axion wave vector and $\omega_a$ its angular frequency which both satisfy the following dispersion relation
\begin{equation}
    \left|\vec k \right|^2 + \frac{c^2 m^2_a}{\hbar^2} = \frac{\omega_a^2}{c^2} \, .
\end{equation}
If the axion field is identified as Dark Matter (DM), its amplitude is then directly related to the local DM density $\rho_\mathrm{DM}$ through\footnote{See \cite{hees18} for a derivation, with the substitution $\varphi \to a/\sqrt{2}$.}
\begin{equation}
    a_0 = \frac{\sqrt{16\pi G \rho_\mathrm{DM}}}{\omega_a c} \, ,
\end{equation}
and the typical amplitude of the wave vector is 
\begin{equation}\label{eq:k}
    k=\left| \vec k \right| \sim \frac{v_\mathrm{DM}\omega_a}{c^2}\sim 10^{-3}\frac{\omega_a}{c} \, ,
\end{equation}
where $v_\mathrm{DM}$ is the typical DM velocity in the Solar System frame. Note that in reality, the DM follows a velocity distribution in our Galaxy such that this amplitude should be treated as a stochastic process. We will discuss this in the following sections.

\subsubsection{Solution including matter}
If we now consider the presence of matter,  Eq.~\eqref{eq:field_eq} corresponds to a modified Klein-Gordon equation for the axion where compact objects (such as Earth) act as an external source. In \cite{hees18}, it has been shown that this source term impacts strongly the amplitude of the oscillation of the free solution from Eq.~\eqref{eq:free_sol}. In particular, depending on the sign of the source term, it can either lead to a screening mechanism, i.e. the amplitude of oscillation is strongly reduced close to the body or to scalarisation, i.e. the amplitude of oscillation is strongly enhanced. The solution derived in \cite{hees18} assumed spherical symmetry and therefore neglects the impact from the $\vec k$ in Eq.~\eqref{eq:free_sol}. Recently, \cite{Banerjee25,DelCastillo25} extended this result by considering non spherical situations, which allows them to consider the impact of the small non vanishing ALP wave vector $\left| \vec k \right|$. In this section, we will summarize the results from \cite{Banerjee25,DelCastillo25} (which assumed flat background spacetime) in the case where the central body is Earth. 

Outside the Earth, the axion solution to Eq.~\eqref{eq:field_eq} with \eqref{eq:sigma}, considering the Earth as a source term modeled as perfect sphere of mass $M_E$ and radius $R_E$  (such that the energy density is $\rho_E = 3M_E c^2/(4\pi R^3_E)$) at location $\vec r$, is the asymptotic DM plane wave from Eq.~\eqref{eq:free_sol} \textit{plus} an additional component which corresponds to circular waves scattered off Earth. The full solution reads (see Eqs. (12)-(16) from \cite{Banerjee25})
\begin{subequations}\label{eq:axion_full_solution}
\begin{align}
    a(t,\vec r) &= \Re \Big[a_{\omega_a}(r,\theta) e^{i(\omega_a t+\Phi)}\Big] = |a_{\omega_a}(r,\theta)|\cos(\omega_a t+\Phi'(r,\theta)) \equiv a_0 A(r,\theta)\cos(\omega_a t+\Phi'(r,\theta)) \, ,
\end{align}
with $\theta$ defined by $\vec k \cdot \vec r = kr\cos \theta$, $\Phi'=\Phi-\arctan(\Im(a_{\omega_a}(r,\theta))/\Re(a_{\omega_a}(r,\theta)))$ and 
\begin{align}\label{eq:axion_multipoles}
 a_{\omega_a}(r,\theta) &= a_0\sum^\infty_{\ell=0}\frac{i^\ell(2\ell+1)P_\ell(\cos\theta)}{Q_\ell}\left(y_\ell(\delta)\Im (Q_\ell)+j_\ell(\delta)\Re (Q_\ell)\right) \, ,
\end{align}
with 
\begin{align}
Q_\ell &= \frac{i}{R_E}\left(\lambda j_{\ell+1}(\lambda)h^{(1)}_\ell(\beta)-\beta j_{\ell}(\lambda)h^{(1)}_{\ell+1}(\beta)\right) \, ,
\end{align}
\end{subequations}
with 
\begin{subequations}\label{eq:lam_del_bet}
\begin{align}
    \delta &= kr \, , \\
    \beta &= kR_E \, , \\ 
    \lambda&=R_E\sqrt{k^2+\kappa \chi_E}=\sqrt{\beta^2-12\frac{E_P^2}{f_a^2}Q_M^E \frac{GM_E}{R_Ec^2}} \, , \label{eq:lambda}
\end{align}
\end{subequations}
where $\chi_E$ is defined in Eq.~(\ref{eq:sigma}), $j_\ell, y_\ell, h^{(1)}_{\ell}$ are respectively the spherical Bessel function of the first kind of order $\ell$, spherical Bessel function of the second kind of order $\ell$ and spherical Hankel function of the first kind of order $\ell$.

The MICROSCOPE experiment is sensitive to low axion mass, typically $m_ac^2 < 10^{-13}$ eV, and is located at $r=R_\mathrm{MIC}\sim R_E + 710$ km \cite{Microscope17}, for which 
\begin{equation}\label{eq:approx_MIC}
kr \leq 4 \times 10^{-3} \ll 1 \, .
\end{equation}
Since a pole $\ell$ of the sum Eq.~\eqref{eq:axion_multipoles} is suppressed by a factor $(kr)^{\ell}$, only the first two contribute, namely the monopole $\ell=0$ and the dipole $\ell=1$. In practice, we find that the monopole is largely dominant over the whole parameter space of interest for MICROSCOPE\footnote{As pointed out in \cite{Banerjee25}, there are some specific couplings such that resonances of the dipole can occur making it the dominant contribution over the monopole. However, in practice, the extremely small frequency/mass width of those resonances (which are $\delta m_a/m_a \sim 10^{-3}\times (m_a c^2/10^{-13} \mathrm{\: eV}) \ll 1$, see \cite{Banerjee25}) make them unimportant when dealing with real data. In addition, our signal being proportional to $a\vec \nabla a$ (see below), the effect of the dipole becomes even less important.}.
Therefore, we use the following expression for the axion field
\begin{align}\label{eq:axion_sol_SM}
    A(r) = \frac{|a_{\omega_a}(r)|}{a_0} &= \frac{\left|\lambda\cos\lambda\sin(\delta-\beta)+\beta\cos(\delta-\beta)\sin\lambda\right|}{\delta\sqrt{\lambda^2\cos^2\lambda+\beta^2\sin^2\lambda}} \, ,
\end{align}
where only $\delta$ depends on $r$, see Eqs.~(\ref{eq:lam_del_bet}).
Taking the $k \rightarrow 0$ limit, $\delta=\beta=0$ and we recover the results derived in \cite{hees18} from the last equation, as expected. 

\begin{figure}
    \centering
    \includegraphics[width=0.45\textwidth]{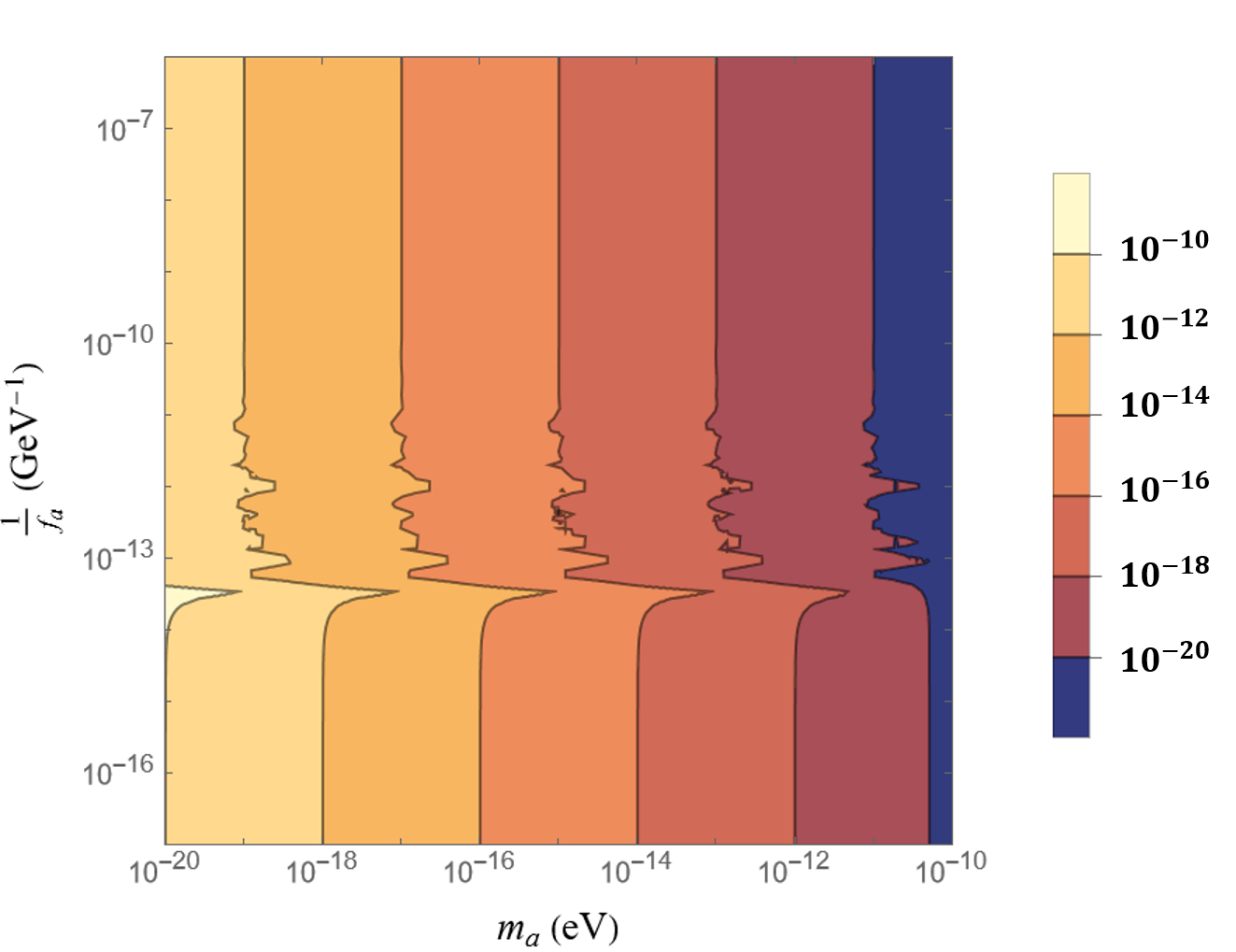}
    \includegraphics[width=0.45\textwidth]{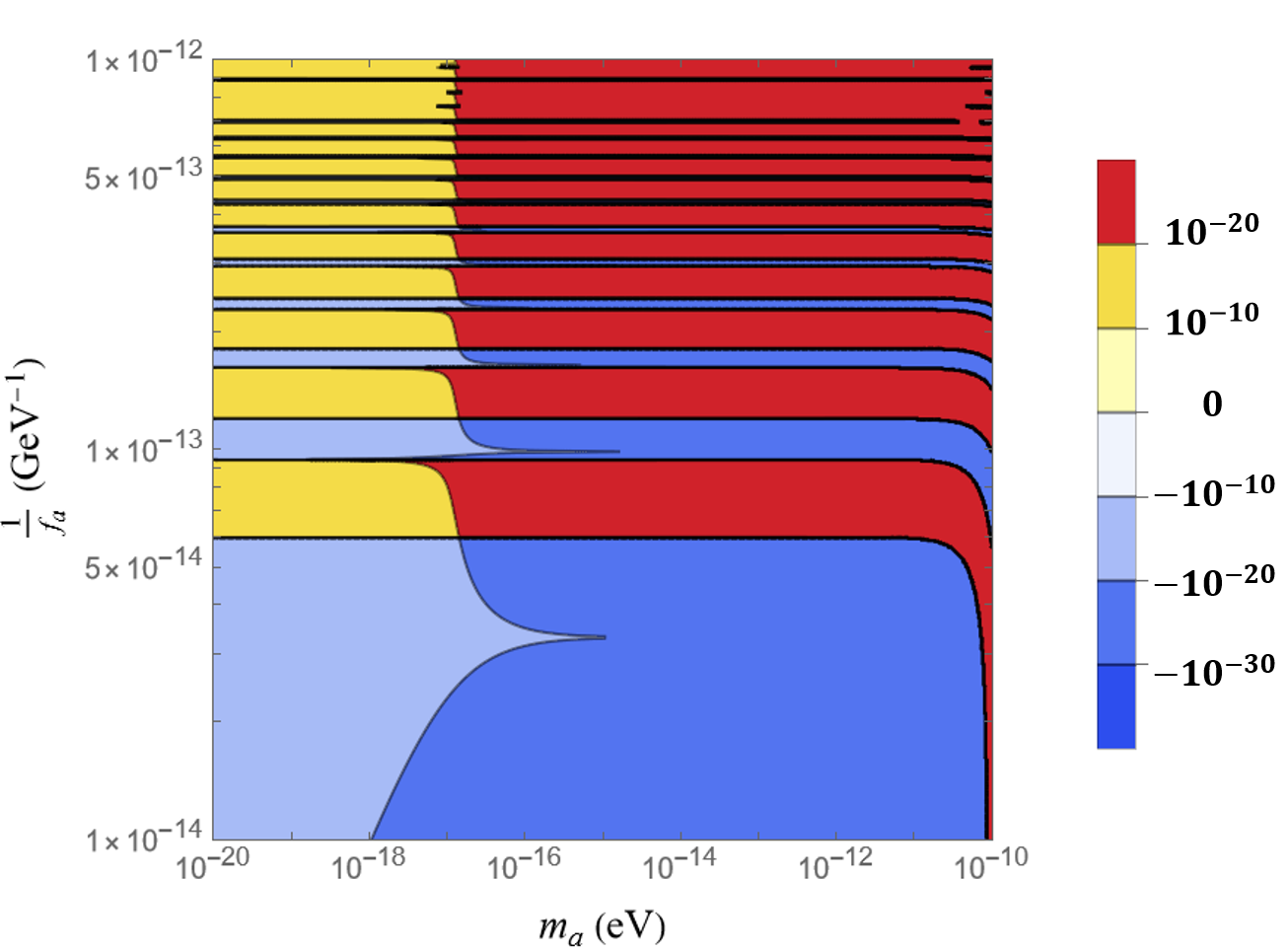}
    \caption{Left panel: evolution of the amplitude of the axion field (in logarithmic scale) $a_0 A(r=R_\mathrm{MIC})$ as a function of $1/f_a$ and of the axion mass $m_a$ in logarithmic scale. The vertical lines indicate the usual $m_a$-dependence of the field amplitude through $a_0 \propto 1/\sqrt{m_a}$, while the horizontal fluctuations are due to resonances from $A(r)$, as discussed in \cite{Banerjee25,DelCastillo25}.
    Right panel: evolution of the amplitude of the radial derivative of the axion field (in logarithmic scale) $a_0 \partial_r A(r=R_\mathrm{MIC})$ as a function of $1/f_a$ and of the axion mass $m_a$. One can notice several sign changes of the gradient. This behavior has important consequence on the derivation of an experimental constraint (see text.). In these figures, the central body is Earth and the axion wave-vector $\left|\vec k\right|=\omega_a v_\mathrm{DM}/c^2$ with $v_\mathrm{DM}/c\sim 10^{-3}$.}
    \label{fig:axion}
\end{figure}

The left panel from Fig.~\ref{fig:axion} shows the evolution of $a_0 A(r=R_\mathrm{MIC})$ in the case where we consider the central body to be the Earth as a function of $1/f_a$ and of the mass of the axion field $m_a$.  As noticed in \cite{Banerjee25,DelCastillo25}, the axion field experienced resonances, i.e. for some values of $1/f_a$, the axion field is strongly enhanced, which is due to the scalarization \cite{hees18}.
The right panel from Fig.~\ref{fig:axion} shows the evolution of the radial gradient of the field $a_0 \partial_r A(r=R_\mathrm{MIC})$, as a function of $1/f_a$ and of the mass of the axion field $m_a$. As it can be noticed from the figure, the gradient changes sign in some specific regions of the parameter space, specifically between $1/f_a\sim 5 \times 10^{-14}$ and $1/f_a\sim 10^{-12}$ GeV$^{-1}$. In particular, it starts negative for very small values of couplings, then, there is a transition where multiple sign changes occur, and finally the gradient becomes positive at large couplings. This feature is very important in deriving constraints because the signal we are looking for is proportional to the field gradient such that there are specific regions for which the signal is theoretically 0 even with finite value of couplings. We will discuss this point more deeply in the final part of this document.

\subsubsection{Dark Matter stochastic distribution}\label{sec:DM_dist}
In a solar system frame, galactic DM follows a velocity distribution centered around $v_\mathrm{DM}\sim$ 230 km/s and with a velocity width $\sigma_v\sim$ 150 km/s, see e.g. \cite{foster:2018aa,Derevianko18,Evans19}
For this reason, the DM axionic field is a stochastic quantity. This has two implications.

First of all, the axion field can be considered as  monochromatic only when the time of observation $T_\mathrm{obs} \ll \tau_\varphi$  where $\tau_\varphi$ is the coherence time of the scalar field, given by \cite{Derevianko18}
\begin{equation}\label{eq:tau_axion}
	\tau_\varphi = \frac{2\pi 10^6}{\omega_a} \, .
\end{equation}

Second, as pointed out in \cite{Centers21,foster:2018aa, Savalle21}, the amplitude of the DM field and its phase are both stochastic variables, even in the monochromatic case when $T_\mathrm{obs} \ll \tau_\varphi$. 

If we restrict ourselves to the case where $T_\mathrm{obs} \ll \tau_\varphi$, the phase $\Phi$ becomes a random uniformly distributed variable while the amplitude follows a Rayleigh distribution \cite{Centers21,foster:2018aa}. For this reason, we write the monochromatic DM axion field as
\begin{subequations}\label{eq:axion_sol_DM}
\begin{equation}
a(t,r)= \alpha \frac{\sqrt{16\pi G \rho_\mathrm{DM}}}{\omega_a c} \frac{\left|\lambda\cos\lambda\sin(\delta-\beta)+\beta\cos(\delta-\beta)\sin\lambda\right|}{\delta\sqrt{\lambda^2\cos^2\lambda+\beta^2\sin^2\lambda}}\cos(\omega_a t +\Phi') \, ,
\end{equation}
where $\alpha$ follows a Rayleigh distribution
\begin{equation}
	P\left[\alpha\right] = \alpha e^{-\alpha^2/2} \, ,
\end{equation}
\end{subequations}
and $\lambda, \beta$ and $\delta$ are provided by Eqs.~(\ref{eq:lam_del_bet}). Note that since $k=\omega_a v_\mathrm{DM}/c^2$, see Eq.~(\ref{eq:k}), it is also a stochastic variable. In our analysis, we neglect its probability distribution function (PDF) because (i) the dependence on $k$ of the axion field amplitude arise through terms like $kr, kR_E \ll 1$, such that effects of the PDF of $k$ arise at second order; and (ii) while the signal we are looking for involves a radial derivative of the field (see next sub-section), one can show it becomes linearly dependent on $k$ only on resonances, which, as mentioned previously are extremely narrow such that their effects are negligible.

\subsection{Acceleration of a test mass}
We now wish to know how the form of the axion field Eq.~\eqref{eq:axion_sol_DM} impacts the acceleration of test masses. The acceleration of a test mass quadratically depending on a scalar field has been derived in \cite{hees18}. The scalar coupling induces several signatures on the acceleration of a test mass: one static term whose amplitude is proportional to the Newtonian gravitational acceleration and several time-depend terms that are oscillating at twice the frequency of the scalar field with location-dependent amplitude. In this work, we will focus on the first term and neglect the fast oscillating terms.

For this, we start from the macroscopic Lagrangian of a given test mass $A$ with rest mass $m_A$ and velocity $\vec v_A$ (see e.g. \cite{damour:2010zr,hees18,Gue:2024onx})
\begin{align}
    \mathcal{L}_A &= -m_A c^2\left(1- \frac{|\vec v_A|^2}{2c^2}\right) \, .
\end{align}
One can easily derive the Euler-Lagrange equation from this Lagrangian and, using Eq.~\eqref{eq:rest_mass_axion}, show that the axion leads to an additional acceleration on a test mass $A$ which reads
\begin{equation}
    \vec a_{A}^\mathrm{axion} = - 2 Q_M^A \left(\frac{E_P}{f_a}\right)^2 a(t,\vec r)\left[c^2\vec \nabla a(t,\vec r) + \vec v_A \dot a(t,\vec r)\right]\, .
\end{equation}
Using Eq.~\eqref{eq:axion_full_solution} leads to
\begin{equation}
    \vec a_{A}^\mathrm{axion} = - 2 Q_M^A \left(\frac{E_P}{f_a}\right)^2 \left|a_{\omega_a}\right| \cos \left(\omega_at+\Phi'\right)\Bigg[c^2\vec\nabla \left(|a_{\omega_a}|\right) \cos \left(\omega_at+\Phi'\right) -  |a_{\omega_a}| \left(c^2\vec\nabla \Phi'+\vec v_A \omega_a\right)\sin \left(\omega_at+\Phi'\right)\Bigg]\, ,
\end{equation}
where $a_{\omega_a}(r,\theta)$ is given by Eq.~(\ref{eq:axion_multipoles}). This acceleration contains fast oscillating terms at $2\omega_a$ and one static (position-dependent) term which reads
\begin{equation}
    \vec a^\mathrm{static}_A = - Q_M^A c^2 \alpha^2 \left(\frac{E_P}{f_a}\right)^2 \left|a_{\omega_a}(r,\theta)\right|\vec\nabla \left(|a_{\omega_a}(r,\theta)|\right)\, ,
\end{equation}
where $\alpha^2$ accounts for the stochastic DM distribution discussed in the previous section.

Considering only the monopole contribution from Eq.~\eqref{eq:axion_multipoles}, i.e. Eq.~\eqref{eq:axion_sol_SM}, one finds one time-independent contribution to the acceleration of the body $A$, which reads
\begin{subequations}
\begin{align}
    \vec a^\mathrm{static}_A &= -c^2\alpha^2 Q^A_M\frac{E^2_P}{f^2_a}\frac{8\pi G \rho_\mathrm{DM}}{\omega^2_a c^2}\frac{\partial A^2(r)}{\partial r}  \hat r \, ,
\end{align}
with $Q^A_M$ the axionic charge of the test mass $A$, and where
\begin{align}
   \frac{\partial A^2(r)}{\partial r} &= \frac{2\left|a_{\omega_a}(r,\theta)\right|\vec\nabla \left(|a_{\omega_a}(r,\theta)|\right)}{a_0} \, \\
   &=k(\sin(\delta-\beta)\lambda\cos\lambda+\beta\cos(\delta-\beta)\sin\lambda) \frac{\cos(\delta-\beta)(\delta\lambda\cos\lambda-\beta\sin\lambda)-\sin(\delta-\beta)(\lambda\cos\lambda+\delta\beta\sin\lambda)}{\delta^3(\lambda^2\cos^2\lambda+\beta^2\sin^2\lambda)} \, . \label{eq:partial_A2}
\end{align}
\end{subequations}
In the previous equation, the random variable $\alpha^2$ is exponentially distributed, i.e 
\begin{equation}\label{eq:exp}
	P\left[\alpha^2\right] = \frac{e^{-\alpha^2/2}}{2}\, .
\end{equation}
Then, the differential acceleration between two bodies $A$ and $B$, with different internal composition, located at the same spacetime position, is  
\begin{align}\label{eq:diff_acc}
    \Delta \vec a =\vec a_A - \vec a_B&= -c^2\alpha^2 \Delta Q_M\frac{E^2_P}{f^2_a}\frac{8\pi G \rho_\mathrm{DM}}{\omega^2_a c^2} \frac{\partial A^2(r)}{\partial r} \hat r\, ,
\end{align}
with $\Delta Q_M = Q^A_M - Q^B_M$. 
Noticing that the static  acceleration is along the same direction as the gravitational acceleration, we can define the E\"otvos parameter as \cite{Touboul22}
\begin{align}\label{eq:signal}
\eta &= \frac{2\Delta a}{a_A + a_B} = \alpha^2 \gamma \, ,
\end{align}
where we introduced
\begin{equation}\label{eq:gamma}
    \gamma = \frac{c^2 \Delta Q_M}{\frac{GM_E}{r^2}}\frac{E^2_P}{f^2_a}\frac{8\pi G \rho_\mathrm{DM}}{\omega^2_a c^2} \frac{\partial A^2(r)}{\partial r} \, .
\end{equation}

\begin{figure}
    \centering
    \includegraphics[width=0.65\textwidth]{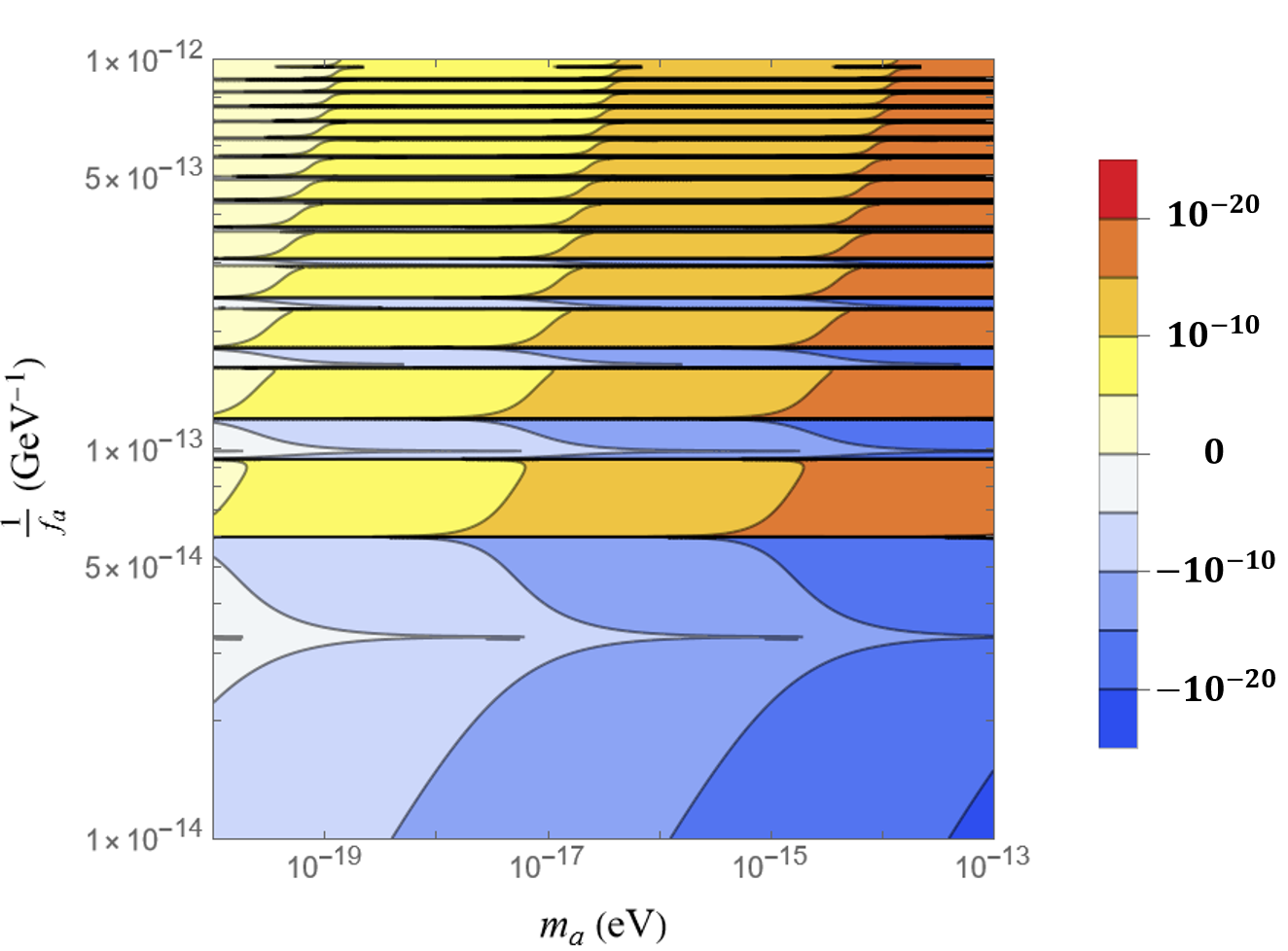}
    \caption{Evolution of $\gamma$ defined by Eq.~(\ref{eq:gamma}) as a function of $1/f_a$ and of the axion mass $m_a$. One can notice clearly multiple sign changes which are due to the evolution of the gradient of the field, see Fig.~\ref{fig:axion}. The resonances from the amplitude of the axion field (see left panel from Fig.~\ref{fig:axion}) and the change of sign of its radial gradient (see right panel from Fig.~\ref{fig:axion}) lead to some bands in $\gamma$. This oscillatory behavior impacts experimental limits on the coupling, see Fig.~\ref{fig:gamma_fa}.}
    \label{fig:gamma}
\end{figure}

Fig.~\ref{fig:gamma} shows the evolution of $\gamma$ as a function of $1/f_a$ and of the axion mass (in the case where the central body is the Earth and the axion wave-vector $k$ is provided by Eq.~(\ref{eq:k})). A 2-dimensional figure for two given axion masses is also provided in Fig.~\ref{fig:gamma_fa}. One can notice that the resonances from the amplitude of the axion field discussed on the left panel of Fig.~\ref{fig:axion} impacts the evolution of $\gamma \propto 1/f_a^2 A(r) \partial_r A(r)$ and leads to the bands visible on this figure.

\subsection{Justification of the Minkowskian background approximation}\label{sec:Mink_approx}
In \cite{hees18}, taking $k=0$, divergences of the field are found when $\lambda$ is an odd multiple of $\pi/2$. In such case, the Minkowskian approximation assumed to derive the axion solution is incorrect and backreactions of the field on the metric must be taken into account. However, in the present case, putting back a non-zero gradient transforms those divergences into resonances which are finite. In this section, we will assess the validity of the Minkowskian approximation used to derive Eq.~\eqref{eq:axion_sol_SM}.

On resonances (i.e. when $\lambda=\pi/2$), the field and its radial gradient become
\begin{subequations}\label{eq:a_resonnance}
\begin{align}
    |a_\omega(r)| &\sim \frac{a_0}{\delta} \, \\
    \partial_r |a_\omega(r)| &\sim \frac{-ka_0}{\delta^2} \, .
\end{align}
\end{subequations}

In using the Minkowski background approximation, we neglect terms that are $\mathcal O\left(h_{\mu\nu}=g_{\mu\nu}-\eta_{\mu\nu}\right)$ in Eq.~\eqref{eq:KG_lin}. The metric field equation from Eq.~\eqref{eq:Einstein_eq} shows that terms in $h_{\mu\nu}$ are generated either by matter or by the axion field. At MICROSCOPE altitude, the terms induced by matter are of the order of $h\sim GM_E/c^2r \sim 6 \times 10^{-10}\ll 1$, which can safely be neglected. We will derive the conditions under which the axion contribution on the right hand side of the field equations from Eq.~\eqref{eq:Einstein_eq} is negligible as well. 

Considering Earth as pressureless matter, i.e $T_{\mu\nu}=\delta_{\mu 0} \delta_{\nu 0} \rho_E$, the Earth contribution to the metric curvature is given by the first two terms of Eq.~\eqref{eq:Einstein_eq}, i.e it is $\kappa \rho_E/2$, at the Minkowskian order. As mentioned in the previous paragraph, the amplitude of the metric perturbation induced by such contribution is $GM_E/c^2r$. Therefore, one can consider that sources of metric perturbations will have an $\mathcal{O}(1)$ contribution if they are of order $\kappa \rho_E/2 \times c^2 r/GM_E$ (which corresponds to a normalization of Earth contribution). Then, the two conditions for the axion to have a negligible contribution to the metric field equation are simply 
\begin{subequations}
\begin{align}
    \frac{rc^2}{GM_E}\frac{\kappa \rho_E}{2} \gg \left(\frac{\partial_t a}{c}\right)^2 - \frac{m^2_a c^2}{\hbar^2}a^2 \, \\
    \frac{rc^2}{GM_E}\frac{\kappa \rho_E}{2} \gg (\partial_r a)^2 + \frac{m^2_a c^2}{\hbar^2}a^2 \, .
\end{align}
\end{subequations}
When evaluated at resonances, i.e. when Eq.~(\ref{eq:a_resonnance}) are fulfilled, these two conditions write
\begin{align}
    \frac{3r^3}{R^3_E} \gg a^2_0 =\frac{16\pi G\rho_\mathrm{DM}}{\omega_a^2 c^2} \quad \mathrm{and} \quad \frac{3r^3}{R^3_E} \gg \frac{a^2_0}{\delta^2}\left(1+\frac{c^2}{v^2_\mathrm{DM}}\delta^2\right) \sim 10^6\frac{16\pi G\rho_\mathrm{DM}}{\omega_a^4 r^2}\left(1+\left(\frac{\omega_a r}{c}\right)^2\right) \, .
\end{align}
Using $r=R_\mathrm{MIC} \sim 7.1 \times 10^6$ m which corresponds to the distance between MICROSCOPE and center of the Earth, we find that the two conditions are fulfilled in the full mass range of interest\footnote{At $m_a c^2 = 10^{-20}$ eV, the LHS of the second equation is larger than the RHS by a factor $\sim 50$, and the ratio increases very quickly at a rate of $m^4_a$.}, such that one can conclude that the Minkowskian approximation is valid.


\subsection{Summary}
To summarize, the axion static contribution to the UFF that is considered in our analysis is provided by Eq.~(\ref{eq:signal}) with $\alpha$ exponentially distributed, see Eq.~(\ref{eq:exp}) and $\gamma$ provided by Eq.~(\ref{eq:gamma}) where the partial derivative of $A^2$ is provided by Eq.~(\ref{eq:partial_A2}). In the expression of $\gamma$, the variables $\beta$, $\lambda$ and $\delta$ are given by Eqs.~(\ref{eq:lam_del_bet}). In particular only $\delta$ depends on the location of the experiment and only $\lambda$ depends on the coupling $1/f_a$ making the relation $\gamma(1/f_a)$ highly non-linear. In all the previous relationships, $k$ is the axion wave-vector which depends on the axion frequency through Eq.~(\ref{eq:k}) and the axion frequency is $\omega_a\approx c^2 m_a/\hbar$. Finally, in the expression of $\gamma$, $\Delta Q_M$ is the axionic charge difference between the two test masses used in MICROSCOPE that are provided in Tab.~\ref{tab:axionic_charge}.


\section{Data analysis using Bayesian inference}\label{app:data_analysis}
In this section, we will present in some detail the methodology used for the data analysis. 

\subsection{MICROSCOPE data}
In this work, we used the published MICROSCOPE data analysis performed sessions by sessions published in \cite{Touboul22cqg} and summarized in Table~\ref{tab:data}. 
We have $N=19$ measurements of the E\"otvos parameter $\eta_i$ and their corresponding statistical and systematics uncertainty. Both these uncertainties are summed quadratically to obtain an estimate of the total uncertainty $\sigma_i$ (last column from Table~\ref{tab:data}) such that the measurements used in this analysis are $\bm \eta=(\eta_1, \dots, \eta_{19})$ and their corresponding total uncertainties $\bm \sigma=(\sigma_1, \dots, \sigma_{19})$ (bold letters denote vectors). 

Each of these estimate of the E\"otvos parameter makes use of MICROSCOPE data taken during one session. In Table~\ref{tab:data}, we also indicate the starting time of each session (given in terms of orbit number, knowing that each orbit lasts for $5946$ s \cite{Touboul22cqg}) as well as their duration. 


\begin{table}
\resizebox{\textwidth}{!}{
\begin{tabular}{|c|c|c|c|c|c|c|}
\hline
Session & Start time & Duration & Measurement & Stat. uncertainty & Systematics & Total uncertainty \\
& (Orbit number) & (s) &($10^{-15}$) & ($10^{-15}$) & ($10^{-15}$) & ($10^{-15}$)\\
\hline
 210  & 4336.5 & 297300 & -29.2 & 13.1 & 1.8 & 13.2\\
 \hline
 212  & 4388 & 452490.6 & 9.5 & 11.9 & 1.0 & 11.9 \\
 \hline
 218  & 4535.1 & 713520 & 6.0 & 8.1 & 1.1 & 8.2\\
 \hline
 234  & 4751.1 & 547032 & 5.9 & 8.3 & 1.0 & 8.4\\
 \hline
 236  & 4844.6 & 713520 & 2.6 & 6.6 & 1.2 & 6.7\\
 \hline
 238  & 4966.1 & 713520 & 5.8 & 6.4 & 1.2 & 6.5\\
 \hline
 252  & 5176.7 & 630276 & -14.9 & 7.3 & 1.1 & 7.4\\
 \hline
 254  & 5284.3 & 713520 & -14.1 & 7.0 & 1.5 & 7.2\\
 \hline
 256  & 5405.8 & 713520 & -5.3 & 7.4 & 1.1 & 7.5\\
 \hline
 326-1  & 7601.5 & 392436 & -16.3 & 9.6 & 1.6 & 9.7\\
 \hline
 326-2  & 7668.5 & 202164 & -10.4 & 13.5 & 1.6 & 13.6\\
 \hline
 358  & 7857.3 & 551788.8 & 15.8 & 10.9 & 1.1 & 11.0\\
 \hline
 402  & 8616.7 & 107028 & 28.4 & 43.6 & 7.3 & 44.2\\
 \hline
 404  & 8637.8 & 713520 & 4.7 & 6.7 & 1.0 & 6.8\\
 \hline
 406  & 8759.3 & 118920 & 5.9 & 14.9 & 3.2 & 15.2\\
 \hline
 438  & 9215.2 & 237840 & -23.4 & 24.6 & 5.5 & 25.2\\
 \hline
 442  & 9298.3 & 237840 & -1.5 & 19.1 & 7.3 & 20.4\\
 \hline
 748  & 12562.3 & 142704 & -23.4 & 24.6 & 7.3 & 25.6\\
 \hline
 750  & 12589.3 & 47568 & 66.9 & 38.4 & 7.3 & 39.1 \\
 \hline
\end{tabular}}
\caption{Data for each session of MICROSCOPE}
\label{tab:data}
\end{table}

\subsection{Likelihood for one data}
 We assume these measurements to be independent and the measurement noise to be distributed following a Gaussian distribution. Using \eqref{eq:signal} to model our signal, the likelihood of one measurement is therefore given by
\begin{equation}\label{eq:individual_like}
    \mathcal L_i(\eta_i,\sigma_i ; \gamma,\alpha) = \frac{1}{\sqrt{2\pi}\sigma_i} e^{-\frac{(\eta_i-\alpha^2\gamma)^2}{2\sigma_i^2}}\, , 
\end{equation}
where $\alpha^2$ follows an exponential distribution (see Eq.~(\ref{eq:exp})) and $\gamma$ is the parameter to be inferred (which is related to the axion coupling, see Eq.~(\ref{eq:gamma})).

\subsection{Creation of groups of measurements spanning one coherence time}\label{sec:group}
We will group these $N$ measurements into $N_g$ subsets, each of them spanning exactly one coherence time of the axion field. This coherence time depends on the axion mass, see Eq.~(\ref{eq:tau_axion}). The reason to perform such a split lies in the fact that the signal we are considering is constant over one coherence time but varies from one coherence time to another. The distribution of the signal over multiple coherence times follows an exponential distribution, as derived in the previous section. Note that the splitting of the $N$ measurements into $N_g$ subsets does not necessarily lead to an even split (in other words: there is not necessarily the same number of measurement in each subset of data).

\subsection{Marginalized likelihood for one subset of data}
Let us consider one subset of data, the $j$th subset (with $1\leq j\leq N_g$. This subset contains $N_j$ measurements (with $1\leq N_j\leq N$): $\bm \eta_j=(\eta_1, \dots, \eta_{N_j})$, $\bm \sigma_j=(\sigma_1, \dots, \sigma_{N_j})$. Since each of these measurements are independent and since the signal is the same for all these measurements (because they are all within one coherence time such that the $\alpha$ parameter that appears in the signal modeling is constant), the likelihood for this subset of data is simply the product of the individual likelihood from Eq.~(\ref{eq:individual_like}). A simple calculation leads to
\begin{subequations}\label{eq:like_group_ind}
\begin{equation}
    \mathcal L_{g_j}(\bm \eta_j, \bm \sigma_j ; \gamma, \alpha) = \prod_{i=1}^{N_j}\mathcal L_i(\eta_i,\sigma_i ; \gamma,\alpha)  = \frac{1}{(2\pi)^{N_j/2} \prod_{i=1}^{N_j}\sigma_i} e^{-\frac{(\bar \eta_j - \gamma\alpha^2)^2}{2\bar \sigma_j^2}} \, ,
\end{equation}
where $\bar \eta_j$ and $\bar \sigma_j$ are the weighted average and uncertainty of the group of measurements that are given by
\begin{align}
    \bar \sigma_j^2 &= \left(\sum_{i=1}^{N_j} \frac{1}{\sigma_i^2}\right)^{-1}\label{eq:bar_sigma}\, , \\
    \bar \eta_j  &= \bar \sigma_j^2 \sum_{i=1}^{N_j} \frac{\eta_i}{\sigma_i^2} \, . \label{eq:bar_eta}
\end{align}
\end{subequations}

Instead of working with a likelihood of the individual data $\bm \eta_j$, we can equivalently use the likelihood of the weighted average, which writes
\begin{equation}\label{eq:like_group}
    \mathcal L_{g_j}(\bar \eta_j, \bar \sigma_j ; \gamma, \alpha)  = \frac{1}{\sqrt{2\pi} \bar \sigma_j} e^{-\frac{(\bar \eta_j - \gamma\alpha^2)^2}{2\bar \sigma_j^2}} \, ,
\end{equation}
which differs from Eq.~(\ref{eq:like_group_ind}) only by a normalization factor.

We can now marginalize this likelihood over the stochastic parameter $\alpha$. The marginalized likelihood is given by
\begin{subequations}\label{eq:marg_like_group}
\begin{align}
		\mathcal L_{\mathrm{marg},g_j}(\bar \eta_j, \bar \sigma_j ;\gamma) &= \int_0^\infty d\left(\alpha^2\right) \mathcal L_{g_j}(\bar \eta_j, \bar \sigma_j ; \gamma, \alpha) P[\alpha^2] =\frac{1}{\sqrt{2\pi} \bar\sigma_j}  \int_0^\infty d\left(\alpha^2\right) e^{-\frac{(\bar \eta_j - \gamma\alpha^2)^2}{2\bar \sigma_j^2}} \frac{e^{-\alpha^2/2}}{2} \nonumber\\
		&=\frac{e^{\frac{\bar \sigma_j^2-4\gamma \bar \eta_j}{8\gamma^2}}}{4\left|\gamma\right|} \mathrm{erfc}\left(\frac{\bar \sigma_j^2-2\gamma \bar\eta_j}{2\sqrt{2}\left|\gamma\right|\bar \sigma_j}\right)  \, , \nonumber\\
		& =\frac{1}{\bar\sigma_j} \tilde {\mathcal L}(\tilde \eta_j, \tilde \gamma_j) \, , \label{eq:marg_like}
\end{align}
where 
\begin{align}
    \tilde{\mathcal L}(\tilde \eta,\tilde \gamma)&=\frac{e^{\frac{1-4\tilde\gamma \tilde \eta}{8\tilde \gamma^2}}}{4\left|\tilde \gamma\right|} \mathrm{erfc}\left(\frac{1-2\tilde\gamma\tilde \eta}{2\sqrt{2}\left|\tilde\gamma\right|}\right) \, , \label{eq:tilde_L}\\
    \tilde \eta_j &=\frac{\bar \eta_j}{\bar \sigma_j}\, ,\\
    \tilde \gamma_j &=\frac{\gamma}{\bar \sigma_j}\, .
\end{align}
\end{subequations}

\begin{figure}
    \centering
    \includegraphics[scale=0.45]{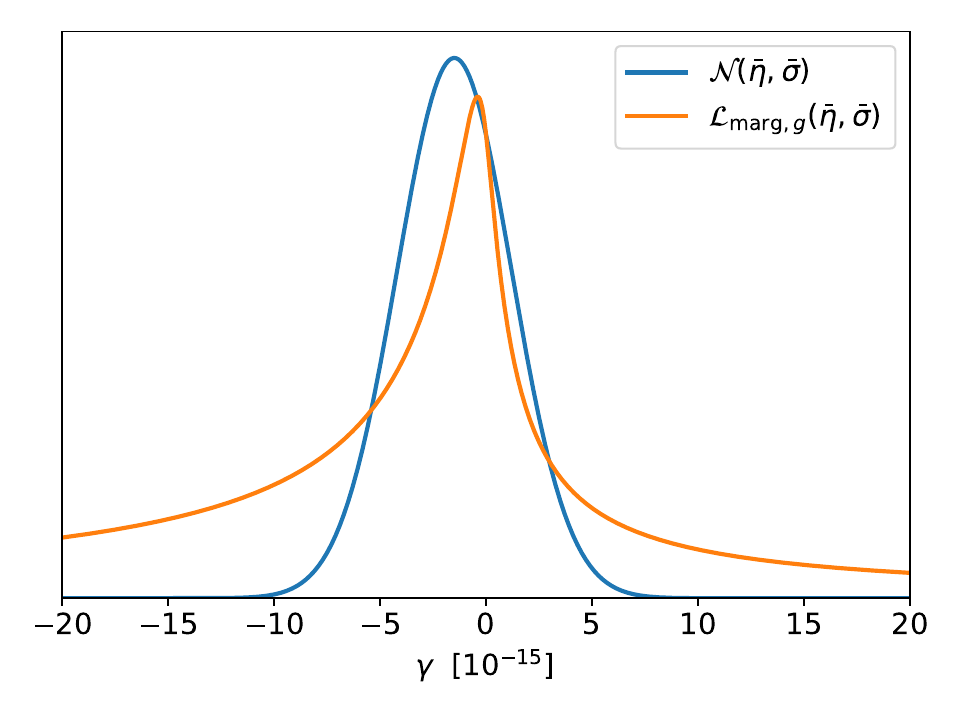}
    \caption{Marginalized likelihood for one subset of data. In blue: the marginalized likelihood assuming a deterministic signal, i.e. using $P[\alpha^2]=\delta\left(\alpha^2-1\right)$ which leads to a normal distribution. In orange: the marginalized likelihood assuming a stochastic signal, given by Eq.~(\ref{eq:marg_like}). For illustrative purposes, we have used values for $\bar \eta$ and $\bar \sigma$ which corresponds to the MICROSCOPE final results, i.e. $\bar \eta=-1.5\times 10^{-15}$ and $\bar \sigma=2.7\times 10^{-15}$, see \cite{Touboul22cqg}.}
    \label{fig:marg_likelihood}
\end{figure}

The marginalized likelihood from Eq.~(\ref{eq:marg_like}) is presented in Fig.~\ref{fig:marg_likelihood} and compared with a normal likelihood, which corresponds to the deterministic case. One can notice that the marginalized likelihood corresponding to the stochastic case is way broader and presents some large tails.

\subsubsection{Various limits of the marginalized likelihood} 
Note that in the limit of $|\tilde\gamma|\rightarrow\infty$, the marginalized likelihood has a $1/|\tilde\gamma|$ asymptotic behaviour since
\begin{equation}\label{eq:barL_asymp}
    \tilde{\mathcal L}(\tilde \eta,\tilde \gamma)\approx \frac{1}{4\left|\tilde \gamma\right|} \mathrm{erfc} \left(-\mathrm{sign} \left(\tilde \gamma\right) \frac{\tilde \eta}{\sqrt{2}}\right) \quad \mathrm{when} \quad |\tilde\gamma|\rightarrow \infty \, .
\end{equation}

In the limit $\gamma \rightarrow 0$, the argument of the error function (erfc) diverges and the function $\tilde{\mathcal L}$ becomes numerically unstable. For such values, it is useful to use the asymptotic expansion of erfc provided by
\begin{equation}\label{eq:erfc_asympt}
	\mathrm{erfc}(z) = \frac{e^{-z^2}}{\sqrt{\pi}z}\left(1-\frac{1}{2z^2}+\frac{3}{4z^4}-\frac{15}{8z^6}+\dots\right)\, ,
\end{equation}	
which leads to an expression for the marginalized likelihood 
\begin{equation}\label{eq:marg_like_small_gamma}
	\tilde{\mathcal L}(\tilde \eta,\tilde \gamma)\approx  \frac{e^{-\frac{\tilde \eta^2}{2}}}{\sqrt{2\pi}\left(1-2\tilde \gamma \tilde \eta\right)} \left[1-4\varepsilon^2+ 48 \varepsilon^4 - 960 \varepsilon^6 +\mathcal O(\varepsilon^8) \right]\, , \quad \mathrm{with} \quad \varepsilon=\frac{\tilde \gamma}{1-2\tilde \gamma \tilde \eta}\, , 
\end{equation}
valid for small $\varepsilon$. 

\subsection{Marginalized likelihood for all the combined subsets of data}
In the previous section, we have derived the marginalized likelihood for one subset of data. In this section, we will combine all the data subsets. Since all the measurements are independent and since the $\alpha_j$ parameter which appears in the fitted model, (i.e. $\gamma \alpha_j^2$) are independent between each subset of data (this is because we have split the data into subset of length equal to the axion coherence time), the total marginalized likelihood is simply given by the product of each individual marginalized likelihood from Eqs.~(\ref{eq:marg_like_group}), i.e.
\begin{subequations}\label{eq:marg_like_tot}
\begin{equation}
    \mathcal L_\mathrm{marg}(\bar{\bm \eta}, \bar {\bm \sigma}, \gamma) = \prod_{j=1}^{N_g}\mathcal L_{\mathrm{marg},g_j}(\bar \eta_j, \bar \sigma_j ;\gamma)=\prod_{j=1}^{N_g} \frac{1}{\bar\sigma_j} \tilde {\mathcal L}\left(\frac{\bar \eta_j}{\bar \sigma_j}, \frac{\gamma}{\bar\sigma_j} \right)\, , 
\end{equation}
\end{subequations}
where $\bar{\bm \eta}=(\bar \eta_1, \dots, \bar \eta_{N_g})$, $\bar{\bm \sigma}=(\bar\sigma_1, \dots, \bar\sigma_{N_g})$ and  $\bar \eta_j$ and $\bar \sigma_j$ are provided in Eqs.~(\ref{eq:like_group}).

\subsection{Prior}
In order to infer a posterior probability distribution on the parameter of interest, i.e. on the axion-gluon coupling $f_a$ which appears in the $\gamma$ parameter through Eq.~(\ref{eq:gamma}), one must specify a prior. In this analysis, following the work from \cite{Centers21}\footnote{In particular, see  Sec. 1.A.2 of the Supplementary Materials of \cite{Centers21}}, we used a Berger–Bernardo reference prior, which is equivalent (since we are working in 1D) to the Jeffrey's prior. This  non informative  prior is motivated because (i) it ensures invariance of the inference results under a change of variables (for example between $f_a$ or $\gamma$), (ii) it ensures that the posterior does not become improper i.e. that its integral does not diverge (which is the case if one chooses naively a uniform prior on $\gamma$), (iii) it maximizes the Kullback–Leibler of the posterior with respect to the prior and (iv) it produces results consistent with the frequentist approach (see e.g. \cite{Centers21}).

The Berger–Bernardo reference prior is equivalent to Jeffrey's prior when working in one dimension and is provided by 
\begin{equation}
	\Pi(\gamma)\propto\sqrt{I(\gamma; \bm \sigma)} \, ,
\end{equation}
where $I$ is the Fisher information that can be computed using
\begin{equation}
	I(\gamma;\bar{ \bm \sigma}) = E\left[\left(\frac{\partial  }{\partial \gamma}\log \mathcal L_\mathrm{marg}(\bar{\bm \eta}, \bar{\bm \sigma}, \gamma)\right)^2 \Big|  \gamma \right]  = -E\left[\frac{\partial^2  }{\partial \gamma^2}\log \mathcal L_\mathrm{marg}(\bar{\bm \eta}, \bar{\bm \sigma}, \gamma)\Big|\gamma \right] \, ,
\end{equation}
which can be written more explicitly as 
\begin{subequations}
\begin{align}
	I(\gamma; \bar{\bm \sigma}) =& \int d\bar{\bm \eta}  \mathcal L_\mathrm{marg}(\bar{\bm \eta}, \bar{\bm \sigma}, \gamma) \left(\frac{\partial  }{\partial \gamma}\log \mathcal L_\mathrm{marg}(\bar{\bm \eta}, \bar{\bm \sigma}, \gamma)\right)^2  \, , \label{eq:Jeffreys_dlog2}\\
	=& -\int d\bar{\bm \eta}  \mathcal L_\mathrm{marg}(\bar{\bm \eta}, \bar{\bm \sigma}, \gamma) \frac{\partial^2  }{\partial \gamma^2}\log \mathcal L_\mathrm{marg}(\bar{\bm \eta}, \bar{\bm \sigma}, \gamma)  \, .\label{eq:Jeffreys_d2log}
\end{align}
Using  Eqs.~(\ref{eq:marg_like_tot}) and~(\ref{eq:Jeffreys_d2log}), we have
\begin{align}
    I(\gamma; \bar{\bm \sigma}) =&- \int d\bar{\bm \eta } \left[\prod_{j=1}^{N_g}\frac{1}{\bar\sigma_j}\tilde {\mathcal L}\left(\frac{\bar \eta_j}{\bar \sigma_j}, \frac{\gamma}{\bar\sigma_j} \right)\right] \left[\sum_{j=1}^{N_g} \frac{\partial^2 }{\partial \gamma^2}\log \tilde {\mathcal L}\left(\frac{\bar \eta_j}{\bar \sigma_j}, \frac{\gamma}{\bar\sigma_j} \right)\right]\,.  
\end{align}
\end{subequations}
Since 
\begin{subequations}
\begin{align}
    \int d\bar \eta_i \Bigg[\frac{1}{\bar\sigma_i}\tilde {\mathcal L}\left(\frac{\bar \eta_i}{\bar \sigma_i}, \frac{\gamma}{\bar\sigma_i} \right) \frac{\partial^2  }{\partial \gamma^2}\log  \tilde {\mathcal L}\left(\frac{\bar \eta_j}{\bar \sigma_j}, \frac{\gamma}{\bar\sigma_j} \right) \Bigg] 
    &= \frac{1}{\bar\sigma_i}\frac{\partial^2  }{\partial \gamma^2}\log \tilde {\mathcal L}\left(\frac{\bar \eta_j}{\bar \sigma_j}, \frac{\gamma}{\bar\sigma_j} \right) \int d\bar\eta_i \tilde {\mathcal L}\left(\frac{\bar \eta_i}{\bar \sigma_i}, \frac{\gamma}{\bar\sigma_i} \right)\\
    &= \frac{\partial^2  }{\partial \gamma^2}\log \tilde {\mathcal L}\left(\frac{\bar \eta_j}{\bar \sigma_j}, \frac{\gamma}{\bar\sigma_j} \right) \, \quad \mathrm{if} \ i\neq j\, , 
\end{align}
\end{subequations}
one can write
\begin{equation}
    I(\gamma; \bar{\bm \sigma}) = \sum_i I_i(\gamma;\bar\sigma_i) 
\end{equation}
with
\begin{align}
   I_i(\gamma;\bar\sigma_i) &= -\frac{1}{\bar\sigma_i}\int_{-\infty}^{\infty} d\bar \eta_i \tilde {\mathcal L}\left(\frac{\bar \eta_i}{\bar \sigma_i}, \frac{\gamma}{\bar\sigma_i} \right) \frac{\partial^2  }{\partial \gamma^2}\log  \tilde {\mathcal L}\left(\frac{\bar \eta_i}{\bar \sigma_i}, \frac{\gamma}{\bar\sigma_i} \right)\nonumber \\
   &= - \frac{1}{\bar\sigma_i^2}\left[\int_{-\infty}^{\infty} dy \tilde {\mathcal L}\left(y,x \right) \frac{\partial^2  }{\partial x^2}\log  \tilde {\mathcal L}\left(y, x \right)\right]_{x=\gamma/\bar\sigma_i} = \frac{\bar I\left(x=\frac{\gamma}{\bar\sigma_i}\right)}{\bar\sigma_i^2} \, ,
\end{align}
where
\begin{equation}\label{eq:barI}
  \bar I\left(x\right) = -\int^\infty_{-\infty} d y   \tilde{\mathcal L}\left(y;x\right) \frac{\partial^2  }{\partial x^2}\log \tilde{\mathcal L}\left(y;x\right) \, .
\end{equation}

As a result, the non informative prior is given by
\begin{equation}\label{eq:prior}
    \Pi(\gamma)= K_\pi \sqrt{\sum_{j=1}^{N_g}\frac{\bar I\left(\frac{\gamma}{\bar\sigma_j}\right)}{\bar\sigma_j^2}} \, ,
\end{equation}
where $K_\pi$ is a normalization factor that ensures that the prior is normalized to unity.

Note that here, we provide an explicit expression of the prior in terms of $\gamma$, one could have equivalently derived the prior on $f_a$. While the form of the prior would be different, the resulting posterior would not be affected since the Jeffrey's prior is invariant under change of variables.

\begin{figure}
    \centering
    \includegraphics[width=0.4\textwidth]{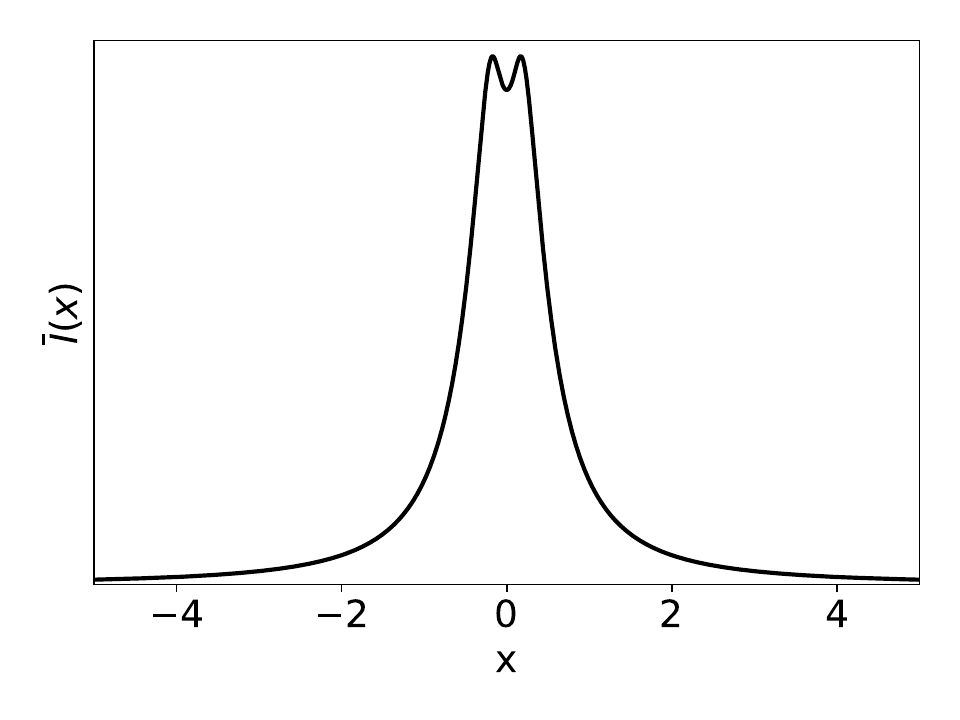}
    \caption{Presentation of the function $\bar I(x)$ defined in Eq.~(\ref{eq:barI}) which characterizes the prior used in our analysis.}
    \label{fig:prior}
\end{figure}

The function $\bar I(x)$ depends only on the function $\tilde {\mathcal L}$ defined in Eq.~(\ref{eq:tilde_L}) and has been computed numerically once. It is presented in Fig.~\ref{fig:prior}.

\subsubsection{Asymptotic behavior of the prior}
The Jeffrey's prior has a simple asymptotic behaviour which can be determined analytically by using Eq.~(\ref{eq:barL_asymp}), which leads to
\begin{equation}
    I_i(\gamma,\bm \sigma) = \frac{1}{\gamma^2} + \mathcal O\left(\frac{1}{\gamma^4}\right) \, ,
\end{equation}
which means that Jeffrey's prior has a $1/\left|\gamma\right|$ behavior for large $\gamma$. Combining this result with the asymptotic expression for the marginalized likelihood, one finds that the product of the likelihood and of the prior has a $1/\gamma^2$ asymptotic shape, which means its integral converges, although a flat prior on $\gamma$ leads to an improper posterior on $\gamma$.

\subsection{Posterior and evidence}
The posterior probability distribution function can be found using Bayes's theorem
\begin{equation}\label{eq:posterior}
	\mathcal P(\gamma ; \bar{\bm \eta}, \bar{\bm\sigma}) = \frac{ \mathcal L_\mathrm{marg}(\bar{\bm \eta}, \bar {\bm \sigma}, \gamma) \Pi(\gamma)}{\mathcal E(\bar{\bm \eta}, \bar {\bm \sigma})} \, , 
\end{equation}
where the marginalized likelihood is provided by Eq.~(\ref{eq:marg_like_tot}), the prior by Eq.~(\ref{eq:prior}) and the evidence is given by
\begin{equation}\label{eq:evidence}
    \mathcal E(\bar{\bm \eta}, \bar {\bm \sigma})= \int_{-\infty}^\infty d\gamma \mathcal L_\mathrm{marg}(\bar{\bm \eta}, \bar {\bm \sigma}, \gamma) \Pi(\gamma) \, .
\end{equation}
This integral is computed numerically.

\begin{figure}
    \centering
    \includegraphics[width=0.8\textwidth]{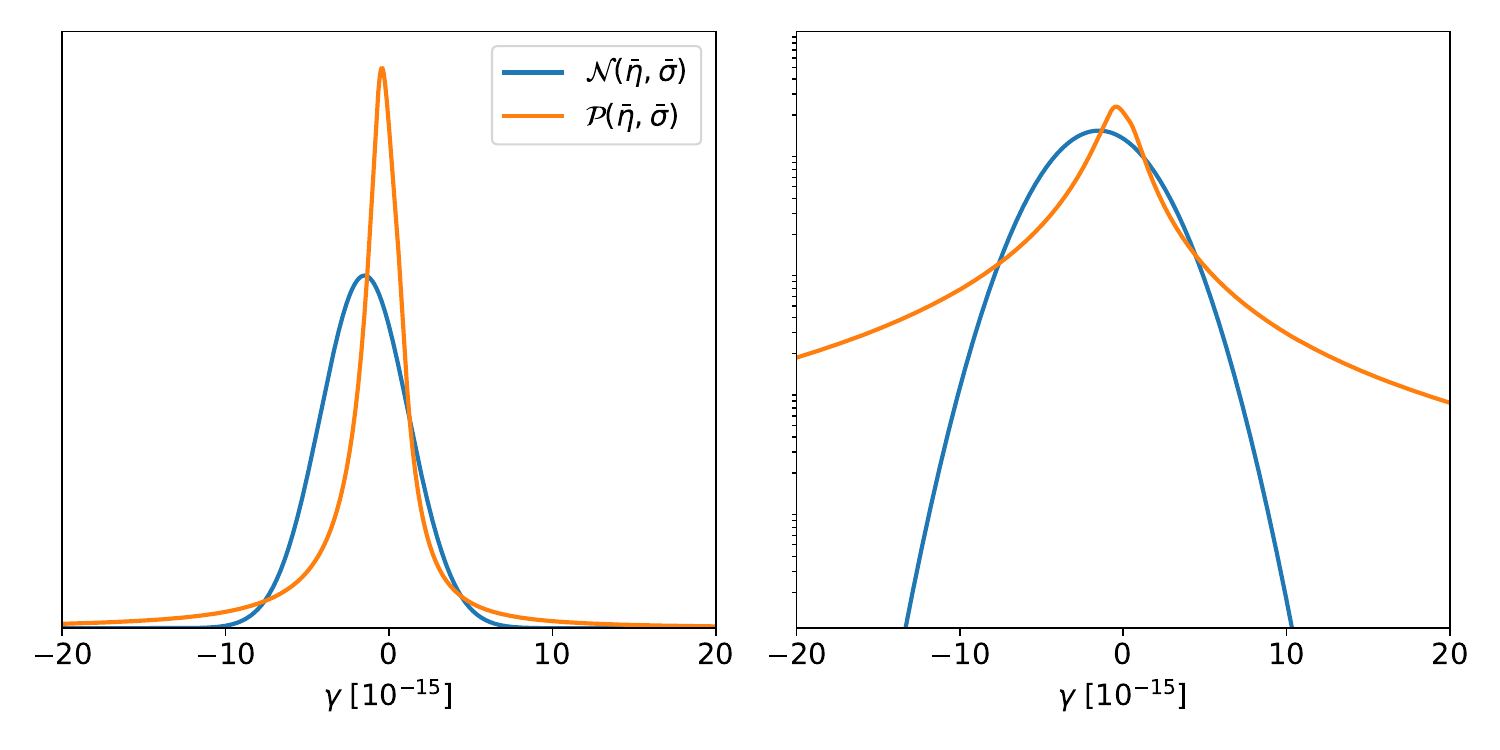}
    \caption{Posterior probability distribution function (left panel: linear scale. right panel: logarithmic vertical scale). In blue: the posterior obtained assuming a deterministic signal, i.e. using $P[\alpha^2]=\delta\left(\alpha^2-1\right)$ which leads to a normal distribution. In orange: the posterior assuming a stochastic signal (i.e. $P[\alpha^2]=e^{-\alpha^2/2}/2$) and using the Jeffrey's prior from Eq.~(\ref{eq:prior}). For illustrative purposes, we have used values for $\bar \eta$ and $\bar \sigma$ which corresponds to the MICROSCOPE final results, i.e. $\bar \eta=-1.5\times 10^{-15}$ and $\bar \sigma=2.7\times 10^{-15}$, see \cite{Touboul22cqg}. The 95\% credible interval for the deterministic posterior is $\left[-6.8\times 10^{-15},3.8\times 10^{-15}\right]$ while it is $\left[-31.6\times 10^{-15},12.6\times 10^{-15}\right]$ in the stochastic case.}
    \label{fig:post}
\end{figure}

Fig.~\ref{fig:post} presents the posterior probability distribution from Eq.~(\ref{eq:posterior}) in the case where one subset of data is used (for illustrative purposes). The blue line presents the posterior if one does not account for the stochasticity of the signal. In this case, both the likelihood and the posterior are normal distribution\footnote{In the case of a normal distribution and a linear model, the Jeffrey's prior is a uniform prior.}. The orange curve from Fig.~\ref{fig:post} presents the posterior if one accounts for the stochasticity of the signal, i.e. using Eq.~(\ref{eq:exp}). In such a case, the marginalized likelihood is provided by Eq.~(\ref{eq:marg_like}) and the Jeffrey's prior by Eq.~(\ref{eq:prior}). The 95\% credible interval is $\sim$ 4 times larger when accounting for the stochasticity of the signal.\footnote{This factor decreases for larger axion masses such that the coherence time is shorter than our total data span. Indeed in that case we use several subsets of data, and the probability that all of them are at a minimum of the axion field amplitude decreases, hence the stochastic ``penalty'' (see e.g. \cite{Centers21}) becomes smaller.} Such a result is similar to the ones obtained in \cite{Centers21} where a linear coupling between matter and the scalar field was explored (while here, we have a quadratic coupling).

\subsection{Bayes factor}
In order to assess if there is a positive detection within the MICROSCOPE data, we use the Bayes factor as a tool for model comparison. More precisely, we compare two models: one with no signal and one with the axion signal from Eq.~(\ref{eq:signal}).

The likelihood corresponding to the case where no signal is fitted to the data is simply given by
\begin{equation}\label{eq:signal_free_like}
    \mathcal L_\mathrm{no\ signal}(\bar{\bm \eta},\bar {\bm \sigma}) = \prod_{j=1}^{N_g}\frac{1}{\sqrt{2\pi} \bar \sigma_j}e^{-\frac{\bar\eta_j^2}{2\bar\sigma_j^2}}\, \, ,
\end{equation}
which corresponds also to the evidence related to the model where no signal is considered. 

\begin{figure}
    \centering
    \includegraphics[width=0.6\textwidth]{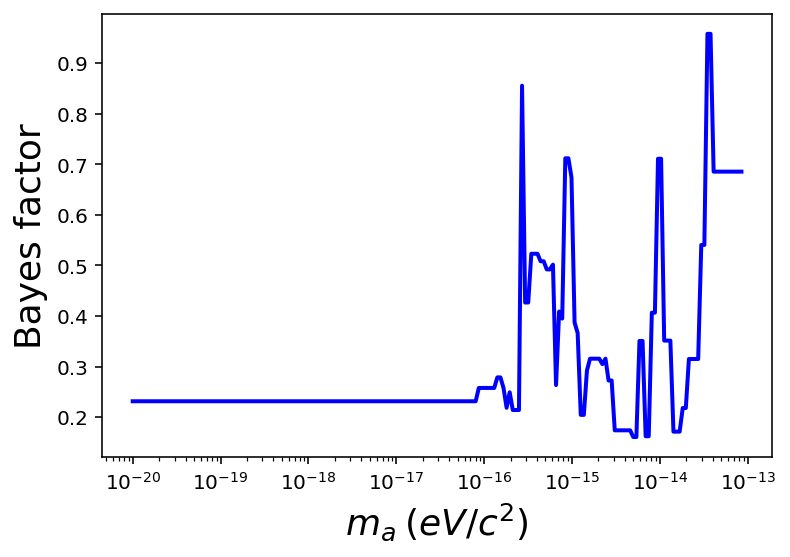}
    \caption{Evolution of the Bayes factor $\mathcal{B}$ defined in Eq.~\eqref{eq:bayes} as function of the axion mass. Over the whole mass range, $\mathcal{B}<1$, indicating no significant detection of axion DM.}
    \label{fig:bayes}
\end{figure}

The Bayes factor in favor of the axion model is computed as the ratio of the evidence of the two models, i.e.
\begin{equation}\label{eq:bayes}
    \mathcal B = \frac{\mathcal E(\bar{\bm \eta}, \bar {\bm \sigma}) }{\mathcal L_\mathrm{no\ signal}(\bar{\bm \eta},\bar {\bm \sigma})} \, ,
\end{equation}
where $\mathcal L_\mathrm{no\ signal}$ is provided by the previous equation and $\mathcal E(\bar{\bm \eta}, \bar {\bm \sigma})$ is provided by Eq.~(\ref{eq:evidence}). This ratio gives the ratio of the probability of having a signal in the data over the probability of having no signal in the data. As pictured in Fig.~\ref{fig:bayes}, the Bayes factor is lower than 1, which is consistent with no axion signal present in the data.

\subsection{Limit on the coupling}\label{sec:constraint_SM}
In the case where the Bayes factor indicates no significant detection in the data, one can infer an upper (or lower) limit on the coupling parameter $1/f_a$ at the $\chi$ upper (lower) limit (here, we will use $\chi=95\%$).

First of all, we compute numerically a lower and upper limit on $\gamma$ by solving
\begin{subequations}
\begin{align}
    \chi = \int_{\gamma^\mathrm{min}_\chi}^{\gamma^\mathrm{max}_\chi} d\gamma \mathcal P(\gamma ; \bar{\bm \eta}, \bar{\bm\sigma}) \, ,
\end{align}
where the posterior is provided by Eq.~(\ref{eq:posterior}). This implicit equation is solved numerically. In practice, we require 
\begin{align}\label{eq:UL_gamma}
    1-\chi &= 2\int_{-\infty}^{\gamma^\mathrm{min}_\chi} d\gamma \mathcal P(\gamma ; \bar{\bm \eta}, \bar{\bm\sigma}) \quad \mathrm{and} \quad 1-\chi = 2\int_{\gamma^\mathrm{max}_\chi}^{\infty} d\gamma \mathcal P(\gamma ; \bar{\bm \eta}, \bar{\bm\sigma}) \, .
\end{align}
\end{subequations}
Note that these limits depend on how the MICROSCOPE sessions are grouped together, which depends on the axion field coherence time and therefore on the axion mass. Put in other words, the limits $\gamma^\mathrm{min}$ and $\gamma^\mathrm{max}$ depend on $m_a$.

\begin{figure}
    \centering
    \includegraphics[width=0.5\textwidth]{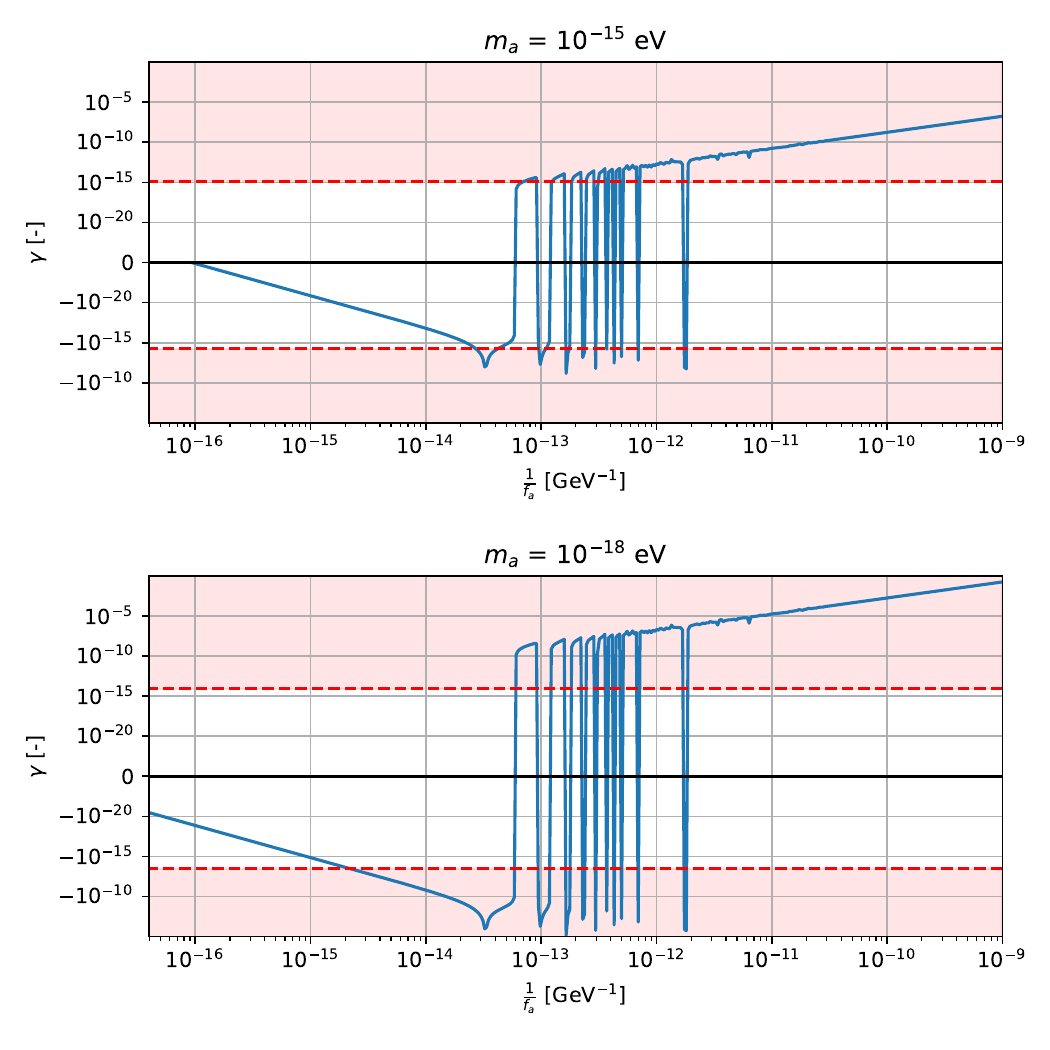}
    \caption{Evolution of $\gamma$ (see Eq.~(\ref{eq:gamma})) as a function of the coupling $1/f_a$ (blue line) for two different values of the axion mass $m_a$, see also Fig.~\ref{fig:gamma}. The dashed red curves correspond to the value of $\gamma^\mathrm{min}$ and $\gamma^\mathrm{max}$ for these two masses obtained from the MICROSCOPE data analysis. The filled area indicates exclusion zone.}
    \label{fig:gamma_fa}
\end{figure}
We then transform these constraints on $\gamma$ on constraints on $1/f_a$ using Eq.~(\ref{eq:gamma}) presented in Fig.~\ref{fig:gamma}, which is also presented in Fig.~\ref{fig:gamma_fa} for two different values of the axion mass. The red dashed curves on \ref{fig:gamma} indicates the value of $\gamma^\mathrm{min}$ and $\gamma^\mathrm{max}$, For a given axion mass $m_a$, we identify the region of the $1/f_a$ parameter space where 
\begin{equation}\label{eq:exclusion_fa}
    \gamma\left(1/f_a\right) < \gamma^\mathrm{min}\, , \qquad \mathrm{or} \qquad \, \gamma\left(1/f_a\right) > \gamma^\mathrm{max}\,  \quad \textrm{for a given } m_a \, ,
\end{equation} 
which is shown by the filled red area on Fig.~\ref{fig:gamma_fa}. Repeating this procedure for different axion masses leads to the exclusion area from our constraint plot presented in the main part of the paper. 

Over the full mass range of interest, $\gamma_\mathrm{min}<0< \gamma_\mathrm{max}$ (in agreement with the fact that our dataset is consistent with no signal). As can be seen from Fig.~\ref{fig:gamma}, in the region $1/f_a < 10^{-13}$ GeV$^{-1}$, $\gamma<0$ the constraint on $1/f_a$ depends on $\gamma_\mathrm{min}$ while for $1/f_a > 10^{-13}$ GeV$^{-1}$, the constraint on $1/f_a$ comes from $\gamma_\mathrm{max}$.

\subsection{Summary}
To summarize, the data analysis is performed as follows for a given mass of the axion field:
\begin{itemize}
    \item compute the coherence time of the axion field using Eq.~(\ref{eq:tau_axion})
    
    \item group the 19 measurements from Table~\ref{tab:data} into subsets of data, where each subset spans exactly one coherence time
    
    \item compute the wighted mean and standard deviation for each subset of data using Eqs.~(\ref{eq:bar_sigma}) and (\ref{eq:bar_eta})
    
    \item compute the evidence of the model that includes a signal using Eq.~(\ref{eq:evidence}) using the marginalized likelihood from Eq.~(\ref{eq:marg_like_tot}) and the prior from Eq.~(\ref{eq:prior}).
    
    \item compute the evidence of the signal-free model using Eq.~(\ref{eq:signal_free_like}).
    
    \item compute the Bayes factor as the ratio of the two evidences computed in the two previous bullet points to assess if a signal is detected or not.
    
    \item in case of no positive detection, determine the 95\% lower and upper limits $\gamma^\mathrm{min}_{0.95}$ and $\gamma^\mathrm{max}_{0.95}$ using Eq.~(\ref{eq:UL_gamma}).
    
    \item transform these limits on $\gamma$ into an exlucion area on $1/f_a$ using Eq.~(\ref{eq:exclusion_fa}) and (\ref{eq:gamma}) presented on Fig.~\ref{fig:gamma}.
\end{itemize}
This procedure is repeated for different axion masses $m_a$.